\documentclass[prb,aps,epsf,floatfix,floats,eqsecnum]{revtex4}
\usepackage{epsfig}
\newcommand{\be}{\begin{equation}}
\newcommand{\ee}{\end{equation}}
\newcommand{\bea}{\begin{eqnarray}}
\newcommand{\eea}{\end{eqnarray}}

\def\nn{\nonumber\\}
\def\r#1{(\ref{#1})}
\def\sgn{{\rm sgn}}
\def\eps{\epsilon}
\begin{document}
%%%%%%%%%%%%%%%%%%%%%%%%%%%%%%%%%%%%%%%%%%%%%%%
\title{Haldane-gap chains in a magnetic field}
%%%%%%%%%%%%%%%%%%%%%%%%%%%%%%%%%%%%%%%%%%%%%%%
\author{Fabian H.L. Essler$^{(a)}$ and Ian Affleck$^{(b)}$}
\affiliation{
$^{(a)}$ The Rudolf Peierls Centre for Theoretical Physics, Oxford
University, Oxford OX1 3NP, UK\\ 
$^{(b)}$ Department of Physics and Astronomy, University of British Columbia, 
  Vancouver, B.C., Canada V6T 1Z1
}
\begin{abstract}
We consider quasi one dimensional spin-1 Heisenberg chains with
crystal field anisotropy in a uniform magnetic field. We determine the
dynamical structure factor in various limits and obtain a fairly
complete qualitative picture of how it changes with the applied
field. In particular, we discuss how the width of the higher
energy single magnon modes depends on the field. We consider the
effects of a weak interchain coupling. We discuss the relevance of our
results for recent neutron scattering experiments on the quasi-1D
Haldane-gap compound NDMAP. 
\end{abstract}

\maketitle
%%%%%%%%%%%%%%%%%%%%%%%%%%
\section{Introduction}
%%%%%%%%%%%%%%%%%%%%%%%%%%
Recent years have seen a resurgence of interest in
field-induced ``magnon condensation'' in gapped quasi one-dimensional
quantum antiferromagnets. In particular a series of ESR and neutron scattering 
experiments have been carried out on the Haldane-gap
\cite{Haldane83} chain compounds NDMAP
\cite{Chen01,az1,az2,az3,Hagiwara03, Tsujii04} 
and NDMAZ\cite{az0}. The main motivation for these experimental studies is
the observation that a spin-1 Heisenberg chain undergoes a quantum
phase transition between a gapped spin-liquid phase and a gapless
Luttinger liquid phase at some critical value $H_c$ of the applied
magnetic field $H$ \cite{Schulz86,ian1}. The ground state of the spin-1
Heisenberg chain is a spin singlet and excitations are described in
terms of a gapped $S=1$ triplet of magnons. When a magnetic field is
applied, the triplet splits due to the Zeeman effect and one of the magnon
gaps is driven to zero at $H_c$. For $H>H_c$ the ground state is magnetized
and excitations are gapless. If interactions between the magnons were
absent, the transition at $H_c$ could be understood as a Bose-Einstein
condensation of magnons. In the spin-1 Heisenberg chain there is an
interaction between magnons, which fundamentally changes the ground
state for $H>H_c$ from a condensate of bosonic magnons to a Luttinger
liquid, which can be regarded as a one dimensional version of an
interacting Bose condensate. The transition at $H=H_c$ is in the
universality class of the 
commensurate-incommensurate (C-IC) phase transition \cite{CIC}.
The magnetic response of the isotropic spin-1 chain in strong
fields $H>H_c$ has been analyzed in some detail in Refs
[\onlinecite{ian2,FZ,KF}]. In Appendix \ref{app:U1} we use the nonlinear
sigma model description of the isotropic spin-1 chain to derive
explicit expressions for the dynamical response functions in the
low-field phase $H<H_c$.
In many $S=1$ compounds such as NENP, NDMAP and NDMAZ strong crystal
field anisotropies are present, which lead to a zero-field splitting
of the magnon triplet comparable in magnitude to the Haldane gap itself.  
These anisotropy effects lead to more complex behavior and a richer
phase diagram. The purpose of this work is to determine the dynamical
response of Haldane-gap compounds in the presence of such crystal
field anisotropies. The relevant lattice Hamiltonian is of the form
\bea
{\cal H}&=&J\sum_j {\bf S}_j\cdot {\bf S}_{j+1}-{\bf H}\cdot {\bf S}_j\nn
&& +\sum_jD (S_j^z)^2+E[(S_j^x)^2-(S_j^y)^2] ,
\label{heisenberg}
\eea
where $0<-E<D$. In zero field there are three magnon modes in the
vicinity of the antiferromagnetic wave number $q=\frac{\pi}{a_0}$ with
gaps $\Delta_1< \Delta_2< \Delta_3$. For simplicity we will mainly
concentrate on the case where the magnetic field is applied along the
z-axis, i.e.  
\be
{\bf H}=H \vec{e}_z.
\ee

The lattice Hamiltonian \r{heisenberg} (with field applied along the
z-direction) exhibits two discrete symmetries which play an important
role in the following: 
\begin{itemize}
\item[1.] Rotation by $\pi$ around the $z$-axis (${\bf R}^z_\pi$):
\bea
S_j^x\rightarrow -S_j^x\ ,\quad
S_j^y\rightarrow -S_j^y\ ,\quad
S_j^z\rightarrow S_j^z.
\label{rzpi}
\eea

\item[2.] Translation by one site (${\bf T}_R$):
\bea
S_j^\alpha&\longrightarrow& S_{j+1}^\alpha\ ,\quad \alpha=x,y,z\ .
\label{tr}
\eea
\end{itemize}

The model \r{heisenberg} is difficult to analyze directly by
analytical methods. However, progress can be made by concentrating on
the low-energy regime, which can be studied by means of
semi-phenomenological descriptions in terms of continuum models. Two
such models have been used in particular, namely (i) a
bosonic Landau-Ginzburg model \cite{ian1,ian2} and 
(ii) a theory of three coupled Majorana fermions \cite{amt}. In the
present work we go beyond the original works \cite{ian1,ian2,amt} by
(1) discussing the decay of high energy magnon modes, (2) applying
methods of integrable quantum field theory to the discussion of
structure functions and (3) taking into account the effects of
inter-chain interactions. 

The outline of this paper is as follows. We
start by reviewing known results on the spectrum of the Majorana
fermion model in section \ref{ssec:Maj1}. We then investigate the role
played by interactions in sections \ref{ssec:Maj2} and \ref{ssec:Maj3} and in
particular show that interactions generate a finite lifetime for one of the
magnon modes. In section \ref{sec:LG} we derive analogous results in
the framework of the Landau-Ginzburg theory. We then turn to a more
detailed analysis of the magnetic response functions in the low-energy
regime in the vicinity of the quantum critical point at $H_c$ in
section \ref{vicinity}. Section \ref{highfield} gives a detailed account
of the low-energy regime in the high-field phase for small
crystal-field anisotropy. The effects of interchain coupling are
investigated in section \ref{interchain} and a summary and discussion of our
various results is given in section \ref{summary}. A variety of technical
details are presented in Appendices \ref{app:U1}-\ref{sineG}.

%%%%%%%%%%%%%%%%%%%%%%%%%%%%%%%%%%%%%%%%%%%%%%%%%%%%%%%%%%%%
\section{Majorana fermion model}
\label{sec:Maj}
%%%%%%%%%%%%%%%%%%%%%%%%%%%%%%%%%%%%%%%%%%%%%%%%%%%%%%%%%%%%
In Ref.[\onlinecite{amt}] A.M. Tsvelik proposed a description of the
spin-1 Heisenberg chain in terms of a field theory of three right and
left moving Majorana (real) fermions $R_a=R_a^\dagger$,
$L_a=L_a^\dagger$, $a=1,2,3$.  The Hamiltonian is given by
\bea
{\cal H}&=&\frac{i}{2}\sum_{a=1}^3 v[L_a\partial_xL_a- R_a\partial_xR_a]
-\Delta_a[R_aL_a-L_aR_a]\nn
&&+\frac{i}{2}\sum_{a,b,c}\varepsilon^{abc}H_a(L_bL_c+R_bR_c)+\ {\cal H}^\prime\ , 
\label{HMajo}
\eea
where ${\cal H}^\prime$ describes anisotropic current-current
interactions  
\bea
{\cal H}^\prime&=&\sum_ag_a J^aJ^a\ .
\label{Hprime}
\eea
Here the currents are bilinears in the Majorana fermions 
\bea
J^a(x)=-\frac{i}{2}\varepsilon^{abc} [L_bL_c+R_bR_c]\ .
\label{current}
\eea

We will take the magnetic field along the 3-axis and concentrate on
the case
\bea
\Delta_1<\Delta_2<\Delta_3\ .
\eea
The generalization of our results to other situations is
straightforward. 
The smooth and staggered components of the spin operators are
defined by the decomposition 
\be
S^\alpha_j\longrightarrow J^\alpha(x)+(-1)^jn^\alpha(x)\ ,
\label{decomposition}
\ee
where $x=ja_0$. The smooth components are equal to the currents,
where $a=1,2,3$ correspond to $\alpha=x,y,z$.
The staggered components are expressed in terms of Ising order and
disorder operators
\bea
n^x(x)\propto\sigma^1(x)\mu^2(x)\mu^3(x)\ ,\nn
n^y(x)\propto\mu^1(x)\sigma^2(x)\mu^3(x)\ ,\nn
n^z(x)\propto\mu^1(x)\mu^2(x)\sigma^3(x)\ .
\label{staggered}
\eea
We note that the signs of the mass terms in \r{HMajo} are such that
in zero field
\be
\langle\sigma^a(x)\rangle =0\ ,\quad 
\langle\mu^a(x)\rangle \neq 0\ .
\ee
%%%%%%%%%%%%%%%%%%%%%%%%%%
\subsection{Symmetries}
%%%%%%%%%%%%%%%%%%%%%%%%%%
The Hamiltonian \r{HMajo} inherits the discrete symmetries
${\bf T}_R$ and ${\bf R}^z_\pi$ from the lattice model. The latter is
realized as 
\bea
{\bf R}^z_\pi:\quad
R_a&\longrightarrow& -R_a\ ,\quad
L_a\longrightarrow -L_a\ ,\nn
\sigma^a&\longrightarrow& -\sigma^a\ ,\quad
\mu^a\longrightarrow \mu^a\ ,\ a=1,2.
\label{MZ2a}
\eea

The translation symmetry by one site turns into a discrete $Z_2$
symmetry in the continuum limit
\bea
J^a(x)\longrightarrow J^a(x)\ ,\qquad
n^a(x)\longrightarrow -n^a(x)\ ,
\eea
and may be realized as
\bea
R_a&\longrightarrow& -R_a\ ,\qquad
L_a\longrightarrow -L_a\ ,\nn
\sigma^a&\longrightarrow&-\sigma^a\ ,\qquad
\mu^a\longrightarrow\mu^a\ ,\ a=1,2,3.
\eea
It is convenient to combine this symmetry with ${\bf R}^z_\pi$ into 
the following $Z_2$ symmetry 
\bea
{\bf T}_R{\bf R}^z_\pi:\quad
R_3&\longrightarrow& -R_3\ ,\quad
L_3\longrightarrow -L_3\ ,\nn
\sigma^3&\longrightarrow&-\sigma^3\ ,\qquad
\mu^3\longrightarrow\mu^3\ .
\label{MZ2b}
\eea
The full symmetry of the Majorana model \r{HMajo} is thus $Z_2\otimes
Z_2$.  
%%%%%%%%%%%%%%%%%%%%%%%%%%%%%%%%%%%%%%%%%%%%%%%%%%%%%%%%%%%%%%
\subsection{Spectrum in the absence of interactions}
\label{ssec:Maj1}
%%%%%%%%%%%%%%%%%%%%%%%%%%%%%%%%%%%%%%%%%%%%%%%%%%%%%%%%%%%%%%
In the absence of interactions, i.e. $g_a=0$ in \r{Hprime} the
Hamiltonian can be diagonalized by means of a Bogoliubov
transformation \cite{amt}. In the following we review some relevant
formulas. As the magnetic field is along the 3-direction only the
first and second Majoranas couple to the magnetic field. 
The third Majorana gives rise to a fermionic single-particle mode with
dispersion 
\bea
\omega_3(k)=\sqrt{\Delta_3^2+v^2k^2}\ .
\label{o3}
\eea
The first and second Majoranas are conveniently combined into a complex
fermion $\Psi$  
\bea
R_1&=&\frac{\Psi_R+\Psi_R^\dagger}{\sqrt{2}}\ ,\quad
R_2=\frac{\Psi_R-\Psi_R^\dagger}{i\sqrt{2}}\ ,\nn
L_1&=&\frac{\Psi_L+\Psi_L^\dagger}{\sqrt{2}}\ ,\quad
L_2=\frac{\Psi_L-\Psi_L^\dagger}{i\sqrt{2}}\ .
\eea
The Hamiltonian density describing the first and second Majoranas takes
the form 
\bea
{\cal H}_{\rm 12}&=&
-iv\left[\Psi^\dagger_R\partial_x\Psi_R-\Psi^\dagger_L\partial_x\Psi_L\right]\nn
&&+\frac{H}{2}\left[[\Psi^\dagger_R,\Psi_R]+[\Psi^\dagger_L,\Psi_L]\right]\nn
&&-im\left[\Psi^\dagger_R\Psi_L-h.c.\right]
+i\Delta\left[\Psi^\dagger_R\Psi^\dagger_L-h.c.\right],
\label{Heff}
\eea
where
\be
\Delta=\frac{\Delta_2-\Delta_1}{2}\ ,\quad
m=\frac{\Delta_2+\Delta_1}{2}.
\ee

Introducing a mode expansion
\bea
\Psi_R(x)&=&\int_{0}^\infty\frac{dk}{2\pi}\left[e^{ikx}\alpha(k)
+e^{-ikx}\beta^\dagger(k)\right]\ ,\nn
\Psi_L(x)&=&\int_{-\infty}^0\frac{dk}{2\pi}\left[e^{ikx}\alpha(k)
+e^{-ikx}\beta^\dagger(k)\right]\ ,
\label{modes}
\eea
we may express the Hamiltonian density \r{Heff} as
\bea
H_{\rm 12}&=&\int_0^\infty\frac{dk}{2\pi}\sum_{a,b=1}^4\gamma^\dagger_a(k)
M_{ab}\gamma_b(k)\ ,
\eea
where
\bea
&&\gamma_a(k)=(\alpha(k),\alpha^\dagger(-k),\beta(k),\beta^\dagger(-k))_a\ ,\nn
&&M=\left[
\begin{array}{cccc}
vk+H & i\Delta & 0 & -im\\
-i\Delta & -vk-H & im & 0\\
0 & -im & vk-H & i\Delta\\
im & 0 & -i\Delta & -vk+H\\
\end{array}
\right].\quad
\eea

Now we perform a Bogoliubov transformation
with a unitary matrix $U(k)$ 
\bea
\pmatrix{c_+(k)\cr c_+^\dagger(-k)\cr c_-(k)\cr c_-^\dagger(-k)\cr}&=&
U(k)\
\pmatrix{\alpha(k)\cr \alpha^\dagger(-k)\cr \beta(k)\cr
  \beta^\dagger(-k)\cr}\ ,
\label{bogo}
\eea
to bring $H_{\rm 12}$ to a diagonal (normal ordered) form
\bea
H_{\rm 12}&=&\int_{-\infty}^\infty\frac{dk}{2\pi}\sum_{a=\pm}
\omega_a(k)\ c^\dagger_a(k)c_a(k)\ ,\nn
\omega_\pm(k)&=&\Bigl[m^2+\Delta^2+H^2+v^2k^2\nn
&&\pm 2\sqrt{m^2\Delta^2+H^2(m^2+v^2k^2)}\Bigr]^\frac{1}{2}.
\label{opm}
\eea
The gap $\omega_+(0)$ increases monotonically with $H$, whereas
$\omega_-(0)$ decreases  and vanishes at a critical field
\be
H_c=\sqrt{\Delta_1\Delta_2}=\sqrt{m^2-\Delta^2}.
\ee
The corresponding critical point is in the Ising universality class.
For fields $H<H_{c2}$ 
\be
H_{c2}=m\left[\frac{1}{2}+\sqrt{\frac{1}{4}+\frac{\Delta^2}{m^2}}
\right]^\frac{1}{2} 
\label{hc2}
\ee
the minimum of the dispersion $\omega_-(k)$
occurs at $k=0$. In the vicinity of $H_c$ and at small momenta the
dispersion is approximated as 
\bea
\omega_-^2(k)&\approx& \frac{H_c^2}{m^2}(H-H_c)^2\nn
&+&
v^2\left[\frac{\Delta^2}{m^2}+\frac{H_c(1+\frac{\Delta^2}{m^2})}{m^2}
(H_c-H)\right]k^2.
\label{omegahc}
\eea
Hence the gap vanishes linearly with $H-H_c$ in agreement with Ising
critical behaviour
\be
\omega_-(0)\simeq (H-H_c)\sqrt{1-\frac{\Delta^2}{m^2}}.
\ee
In order to see the Ising criticality described by \r{omegahc} the
magnetic field must be sufficiently close to $H_c$
\be
H-H_c<\frac{\Delta^2}{H_c(1+\frac{\Delta^2}{m^2})}.
\label{isingapplicable}
\ee
As expected this scale is set by the anisotropy $\Delta$. For
$H>H_{c2}$ there are two degenerate minima of $\omega_-(k)$ at some
incommensurate wave numbers $\pm k_F$ with 
\bea
vk_F&=&\left[H^2-m^2-\frac{m^2\Delta^2}{H^2}\right]^\frac{1}{2}\ ,\nn
\omega_-(k_F)&=&\Delta\sqrt{1-\frac{m^2}{H^2}}.
\eea
For $k\approx \pm k_F$ we have
\bea
\omega_-^2(k)\approx 
\Delta^2\left[1-\frac{m^2}{H^2}\right]
+\left[\frac{v^2k_F}{H}\right]^2(|k|-k_F)^2.
\label{omegahc2}
\eea
As was recently pointed out by Wang \cite{wang}, these results for the
dispersions suggest that a cross-over between Ising and
C-IC critical behaviour occurs as a function of $H$. For $H$ very
close to $H_c$ we encounter Ising critical behaviour, which crosses
over to C-IC behaviour for $H>H_{c2}$. 

%%%%%%%%%%%%%%%%%%%%%%%%%%%%%%%%%%%%%%%%%%%%%%%%%%%%%%%%%%%%%%%
\subsection{Interactions: self-consistent mean-field treatment}
\label{ssec:Maj2}
%%%%%%%%%%%%%%%%%%%%%%%%%%%%%%%%%%%%%%%%%%%%%%%%%%%%%%%%%%%%%%%
So far we have neglected the four-fermion interactions \r{Hprime}
altogether. As a first step of taking interactions into account we may
treat them in a self-consistent mean-field approximation (SCMF). The
following expectation values are compatible with the discrete $Z_2$
symmetries 
\bea
&&\langle L_aR_a\rangle\neq 0\ ,\
\langle L_1L_2\rangle\neq 0\ ,\
\langle R_1R_2\rangle\neq 0\ ,\nn
&&\langle R_1L_2\rangle\neq 0\ ,\
\langle R_2L_1\rangle\neq 0\ .
\eea
Decoupling the four fermion terms leads to a renormalization of the
mass parameters 
\bea
\Delta_1&\longrightarrow&\tilde\Delta_1=
\Delta_1-2ig_2\langle L_3R_3\rangle-2ig_3\langle L_2R_2\rangle,\nn
\Delta_2&\longrightarrow&\tilde\Delta_2=
\Delta_2-2ig_1\langle L_3R_3\rangle-2ig_3\langle L_1R_1\rangle,\nn
\Delta_3&\longrightarrow&\tilde\Delta_3=
\Delta_3-2ig_3\langle L_2R_2\rangle-2ig_2\langle L_1R_1\rangle\ .
\eea
The magnetic field terms are changed to
\be
H_L\ L_1L_2+ H_R\ R_1R_2\ ,
\ee
where $H_L=H+2ig_3\langle R_1R_2\rangle$. and $H_R=H+2ig_3\langle
L_1L_2\rangle$. Finally, two new terms are generated
\be
i\lambda_1\ R_1L_2+
i\lambda_2\ L_1R_2\ ,
\ee
where $i\lambda_1=2g_3\langle L_1 R_2\rangle\ $ and
$i\lambda_2=2g_3\langle R_1 L_2\rangle$. The resulting mean-field
Hamiltonian is quadratic in the Fermi fields and can again be
diagonalized by a Bogoliubov transformation. Let us denote the ground
state energy obtained in this way by $E_{\rm GS}$.
The expectation values are determined self-consistently, e.g.
\bea
-i\langle R_aL_a\rangle&=&\frac{\partial E_{\rm
GS}}{\partial\tilde{\Delta}_a}\ . 
\eea
The SCMF procedure is easily implemented once the appropriate
couplings $g_a$ are specified. However, in order to keep
matters as simple as possible we will assume from now on that the
$g_a$ are small and as a result the differences between the free
theory and the SCMF theory are negligible. 
The main qualitative effect of nonzero $g_a$ is to make the gap of the
third Majorana magnetic field dependent. Such a dependence is observed
for example in experiments on NDMAP\cite{az2,az3} and implies the
presence of interactions in the framework of the Majorana fermion
model. 

%%%%%%%%%%%%%%%%%%%%%%%%%%%%%%%%%%%%%%%%%%%%%%%%%%%%
\subsection{Decay Processes in the Low-Field Phase}
\label{ssec:Maj3}
%%%%%%%%%%%%%%%%%%%%%%%%%%%%%%%%%%%%%%%%%%%%%%%%%%%%
Within the SCMF approximation the role of the current-current
interactions is merely to induce slight changes of the dispersion
relations of the three coherent single-particle magnon modes.
As we will now show, treating the interactions beyond the SCMF
approximation leads to the damping of one of the magnons.

The analysis of the spectrum summarized above establishes that there
are three different types of magnons, which we will refer to as $M_3$,
$M_+$ and $M_-$ respectively. The corresponding dispersion relations
are $\omega_3(k)$ \r{o3} and $\omega_\pm(k)$ \r{opm}. The interaction
of these modes is described by the term ${\cal H}^\prime$ in the
Hamiltonian \r{HMajo} and involves four particles.
As $\omega_-(k)$ can become very small as the
magnetic field is increased from zero, the decays $M_3\longrightarrow
M_-M_-M_-$ and $M_+\longrightarrow M_-M_-M_-$ become kinematically
allowed for sufficiently large magnetic fields. However, the decay of
$M_3$ is forbidden by the symmetry ${\bf T}_R{\bf R}^z_\pi$: $M_3$ is
odd under this symmetry whereas $M_\pm$ are even. Essentially we are
using the fact that all 3 magnon modes only exist near wave-vector
$\frac{\pi}{a_0}$ so that they can only decay into an odd number of 
magnons.  The decay of $M_3$ into any odd number of $M_-$'s would be
inconsistent with the $R^z_{\pi}$ spin rotation symmetry.
On the other hand the decay $M_+\longrightarrow M_-M_-M_-$ is allowed
by symmetry. The process becomes kinematically possible as soon as the
magnetic field exceeds a critical value $H_{\rm d}$, which is defined by 
\bea
\omega_+(0)=3\omega_-(0)\ .
\eea
Solving for $H_{\rm d}$ we find
\bea
H_{\rm d}&=&\sqrt{\frac{m^2}{4}-\Delta^2}\ .
\eea
As long as $2\Delta<m$ the decay process will occur for $H>H_{\rm
  d}$ and from now on we will assume that this is the case.

We note that in zero field there are no decay processes even if the
gaps are such that they are kinematically allowed
($3\Delta_1<\Delta_2$). The reason is that for $H=0$ the lattice
Hamiltonian \r{heisenberg} has additional spin rotational symmetries
around the $x$ and $y$ axes by 180 degrees. In combination with the
translation symmetry these induce symmetries of the form
\r{MZ2b} for the Majoranas $1$ and $2$ individually rather than the
combination \r{MZ2a}. This additional symmetry forbids decay
processes. 

Inserting the mode expansions \r{modes} into the expression for ${\cal
H}'$, integrating over the spatial coordinate and then
carrying out the Bogoliubov transformation \r{bogo} generates
several terms quartic in the fermionic creation annihilation
operators $c_{\pm}(k)$, $c^\dagger_{\pm}(k)$. The most interesting one
describes the decay of a $M_+$ mode into three $M_-$ modes and is of the form
\bea
V&=&g_3\int_{-\infty}^\infty\frac{dk_1dk_2dk_3}{(2\pi)^3}\ f(k_1,k_2,k_3)\nn
&&\quad\times c^\dagger_-(k_1)c^\dagger_-(k_2)c^\dagger_-(k_3)c_+(k_1+k_2+k_3).
\eea
Here $f$ is an antisymmetric function of its arguments and at small
momenta is of the form
\bea
f(k_1,k_2,k_3)\simeq \tilde{C}\ (k_1-k_2)(k_1-k_3)(k_2-k_3).
\label{fsmallk}
\eea
The constant $\tilde{C}$ is a complicated function of $\Delta$, $m$ and $H$.
The differential rate for the decay of a $M_+$ particle with momentum
$p$ into three $M_-$ particles with momenta $p_1,p_3,p_3$ is
\bea
d\Gamma&=&2\pi|M|^2\delta(p-\sum_{j=1}^3p_j)\
\delta(\omega_+(p)-\sum_{j=1}^3\omega_-(p_j))\nn
&&\times\frac{dp_1dp_2dp_3}{3!}\ ,
\label{dGamma}
\eea
where the factor of $3!$ is introduced to account for the fact that
the three particles in the final state are indistinguishable.
In the Born approximation the transition matrix element $M$ is
\bea
M&=&\frac{1}{4\pi^2}\langle0|\prod_{j=1}^3c_-(p_j)\ V\
c_+^\dagger(p)|0\rangle\Bigl(\delta(p-\sum_{j=1}^3p_j)\Bigr)^{-1}\nn
&=&\frac{g_3\ 3!}{2\pi}\ f(p_1,p_2,p_3).
\eea
Taking the $M_+$ particle to be at rest, i.e. setting $p=0$, we obtain
\bea
\Gamma&=&\frac{6g_3^2}{2\pi}\int dp_1dp_2|f(p_1,p_2,-p_1-p_2)|^2\nn
&\times& \delta(\omega_+(0)-\omega_-(p_1)-\omega_-(p_2)-\omega_-(p_1+p_2)).
\label{decayrate}
\eea
In order to simplify matters further, we concentrate on the case where
the magnetic field is close the critical field $H_{\rm d}$ at which the
decay $M_+\rightarrow M_-M_-M_-$ first becomes kinematically
possible. In this regime we have
\be
\omega_+(0)-3\omega_-(0)\ll \omega_-(0)\ .
\ee
Then the momenta $p_{1,2}$ in
\r{decayrate} have to be small in order to satisfy the delta-function
and we may use the expansions \r{fsmallk} for $f$ and
\bea
\omega_-(p)&\approx& \omega_-(0)+\tilde{\alpha}p^2+{\cal O}(p^4)\ ,\nn
\tilde{\alpha}&=&v^2\left[1-\frac{H^2}{m\sqrt{H^2+\Delta^2}}\right][2\omega_-(0)]^{-1}.
\eea
This leads to the following expression for the decay rate in the
regime $H>H_{\rm d}$, $\frac{H}{H_{\rm d}}-1\ll 1$
\bea
\Gamma&\approx&\frac{6g_3^2}{2\pi}\frac{\tilde{C}^2}{(2\tilde{\alpha})^4}
\frac{4\pi}{\sqrt{3}}\ [\omega_+(0)-3\omega_-(0)]^3\nn
&\approx&g_3^2\sqrt{3}\tilde{C}^2
\left[\frac{2}{\tilde{\alpha}}\right]^4
\left[1-\frac{\Delta^2}{4m^2}\right]^\frac{3}{2}(H-H_{\rm d})^3\ . 
\eea
We find that the decay rate is proportional to $(H-H_{\rm d})^3$ and
is therefore quite small in the vicinity of $H_{\rm d}$.

%%%%%%%%%%%%%%%%%%%%%%%%%%%%%%%%%%%
\section{Landau-Ginzburg (LG) Model}
\label{sec:LG}
%%%%%%%%%%%%%%%%%%%%%%%%%%%%%%%%%%%
A different approach to studying the spin-1 Heisenberg chain with
crystal field anisotropies in a magnetic field was used in Refs
[\onlinecite{ian1,ian2}]. It is based on the nonlinear sigma model
description of the spin-S Heisenberg chain in terms of the
three-component field $\vec\varphi$ describing the staggered
components of the spin operators and the subsequent
approximation of replacing the constraint $\vec{\varphi}^2=1$ by a
$|\vec{\varphi}|^4$ interaction. The LG Lagrangian
density is \cite{ian1,ian2} 
\bea
{\cal L}&=&\frac{1}{2v}\left(
\frac{\partial\vec{\varphi}}{\partial t}+\vec{H}\times\vec{\varphi}\right)^2
-\frac{v}{2}\left(
\frac{\partial\vec{\varphi}}{\partial x}\right)^2\nn
&&-\sum_{a=1}^3\frac{\Delta_a^2}{2v}\varphi^2_a-\lambda|\vec{\varphi}|^4.
\label{LG}
\eea
We again take the magnetic
field to point along the $3$-axis
\be
\vec{H}=H\vec{e}_3.
\ee
It then follows from \r{LG} that $\varphi_3$ couples to the magnetic
field only via the $\lambda|\vec{\varphi}|^4$ interaction. The
Landau-Ginzburg theory inherits the discrete symmetries \r{rzpi}
and \r{tr} from the underlying lattice model. As is the Majorana
fermion model it is convenient to combine the translational symmetry
by one site ${\bf T}_R$ with the rotation around the z-axis by $\pi$ 
${\bf R}^z_\pi$ and obtain the following two $Z_2$ symmetries
\bea
{\bf R}^z_\pi:\quad
\varphi_a\longrightarrow -\varphi_a\ ,\quad\ a=1,2.
\label{LGZ2a}
\eea
\bea
{\bf T}_R{\bf R}^z_\pi:\quad
\varphi_3\longrightarrow -\varphi_3\ .
\label{LGZ2b}
\eea
The resulting $Z_2\otimes Z_2$ symmetry is the same as for the Majorana
fermion model.
%%%%%%%%%%%%%%%%%%%%%%%%%%%%%%%%%%%%%%%%%
\subsection{Spectrum and Mode Expansion}
%%%%%%%%%%%%%%%%%%%%%%%%%%%%%%%%%%%%%%%%%
Neglecting the $|\vec{\varphi}|^4$ interaction the spectrum can be
determined by solving the classical equations of motion
\bea
\left[\frac{\partial^2}{\partial t^2}
-v^2\frac{\partial^2}{\partial x^2}
+\Delta_a^2-H^2\right]\varphi_a
+2\varepsilon^{abc}H_b\frac{\partial\varphi_c}{\partial t}=0.
\label{LGEOM}
\eea
One finds that there are three magnon modes $M_{3}$ and $M_\pm$ with
the following dispersion relations \cite{ian1,ian2}
\bea
&&\omega_3(k)=\sqrt{\Delta_3^2+v^2k^2}\ ,\nn
&&\omega_\pm(k)=\Biggl[H^2+\frac{\Delta_1^2+\Delta_2^2}{2}+v^2k^2\nn
&&\pm\sqrt{2H^2(\Delta_1^2+\Delta_2^2+2v^2k^2)+\left(\frac{\Delta_1^2-\Delta_2^2}{2}
\right)^2}\Biggr]^\frac{1}{2} .\quad
\label{o3pm}
\eea
The low energy $M_-$ mode with dispersion $\omega_-(k)$ becomes
gapless at a critical field $H_c=\Delta_1$. The resulting critical
point is in the universality class of the two-dimensional Ising model
\cite{ian2}. 

The scalar fields $\varphi_{1,2}$ have the following mode expansions
\bea
\varphi_1(t,x)&=&\sum_{\alpha=\pm}\int dk\frac{
A_{1\alpha}(k)e^{-i\omega_\alpha(k)t+ikx}a_\alpha(k)+{\rm h.c.}}
{\sqrt{4\pi\omega_\alpha(k)/v}},\nn
\varphi_2(t,x)&=&\sum_{\alpha=\pm}\int dk\frac{
iA_{2\alpha}(k)e^{-i\omega_\alpha(k)t+ikx}a_\alpha(k)+{\rm h.c.}}
{\sqrt{4\pi\omega_\alpha(k)/v}},\nn
\label{LGmodes}
\eea
where $A^*_{a\alpha}(k)=A_{a\alpha}(k)$.
Here $a$ and $a^\dagger$ obey canonical commutation relations
\bea
[a_\alpha(k),a^\dagger_\beta(p)]=\delta_{\alpha\beta}\ \delta(k-p)\ ,
\eea
and the amplitudes $A_{a\pm}(k)$ are fixed by the requirements that
(i) the fields $\varphi_a$ fulfil the equations of motion \r{LGEOM}
and (ii) the fields $\varphi_a$ and the conjugate momenta $\Pi_a
=\frac{1}{v}\frac{\partial \varphi_a}{\partial
t}+(\vec{H}\times\vec{\varphi})_a$
fulfil canonical commutation relations 
\bea
&&[\varphi_a(t,x),\varphi_b(t,y)]=0\ ,\quad
[\Pi_a(t,x),\Pi_b(t,y)]=0\ ,\nn
&&[\varphi_a(t,x),\Pi_b(t,y)]=i\delta_{ab}\delta(x-y)\ .
\eea
We find
\bea
A_{1+}^2(k)&=&\frac{3H^2+\Delta_1^2+v^2k^2-\omega_-(k)^2}
{\omega_+(k)^2-\omega_-(k)^2}\ ,\nn 
A_{1-}^2(k)&=&1-A_{1+}^2(k)=\frac{\omega_+(k)^2-3H^2-\Delta_1^2-v^2k^2}
{\omega_+(k)^2-\omega_-(k)^2}\ ,\nn 
\frac{A_{2a}(k)}{A_{1a}(k)}&=&-
\frac{\omega_a(k)^2+H^2-\Delta_1^2-v^2k^2}{2H\omega_a(k)}\ ,\ a=\pm.
\label{apm}
\eea
Eqns \r{apm} allow us to deduce the polarizations of the modes
corresponding to the dispersion $\omega_\pm(k)$. For example, at $H=0$
the $M_-$ mode is polarized entirely along the 1 direction and the
$M_+$ mode along the 2 direction. On the other hand, as $H\rightarrow
\Delta_1$ we find that
\bea
A_{2-}(0)&\rightarrow& 0\ ,\quad A_{2+}^2(0)\rightarrow 1\ ,\nn
A_{1-}^2(0)&\rightarrow& \frac{{\Delta_2^2-\Delta_1^2}}
{{3\Delta_1^2+\Delta_2^2}}\ ,\quad 
A_{1+}^2(0)\rightarrow \frac{4\Delta_1^2}{{3\Delta_1^2+\Delta_2^2}}.
\label{apm2}
\eea
Hence $\varphi_2$ couples only to the $M_+$ mode whereas $\varphi_1$
couples to the $M_+$ mode as well as to the $M_-$ mode with a strength
set by the anisotropy.  
%%%%%%%%%%%%%%%%%%%%%%%%%%%%%%%
\subsection{Decay Processes}
%%%%%%%%%%%%%%%%%%%%%%%%%%%%%%%
In the absence of the nonlinear $|\vec{\varphi}|^4$-term
the LG model describes three coherent magnons $M_3$, $M_\pm$ with
corresponding dispersion relations \r{o3pm}. Inclusion of the
$|\vec{\varphi}|^4$ term generates interaction terms involving four
particles. As was the case in the Majorana fermion model,
$\omega_-(k)$ can become very small when the magnetic field is
increased and as a result the decays $M_3\longrightarrow M_-M_-M_-$
and $M_+\longrightarrow M_-M_-M_-$ become kinematically allowed. The
decay of $M_3$ is forbidden by the symmetry ${\bf T}_R{\bf R}^z_\pi$
\r{LGZ2b}: $M_3$ is odd under this symmetry whereas $M_\pm$ are
even. On the other hand the decay process $M_+\rightarrow 3M_-$ is
allowed when the magnetic field is larger than 
\bea
H_{\rm d}&=&\Biggl[\frac{17}{16}[\Delta_1^2+\Delta_2^2]\pm
\frac{5}{16}\sqrt{13[\Delta_1^4+\Delta_2^4]+10\Delta_1^2\Delta_2^2}
\Biggr]^\frac{1}{2}.\nn
\eea
The interaction describing this decay process is given by
\bea
V&=&\frac{\lambda v^2}{2\pi}
\int_{-\infty}^\infty dk_1dk_2dk_3\ g(k_1,k_2,k_3)\nn
&&\quad\times a^\dagger_-(k_1)a^\dagger_-(k_2)a^\dagger_-(k_3)a_+(k_1+k_2+k_3),
\eea
where $g(k_1,k_2,k_3)$
%\begin{widetext}
%\bea
%g(k_1,k_2,k_3)&=&\left[A_{1-}(k_1)A_{1-}(k_2)-A_{2-}(k_1)A_{2-}(k_2)\right]
%\left[A_{1-}(k_3)A_{1+}(k_1+k_2+k_3)+A_{2-}(k_3)A_{2+}(k_1+k_2+k_3)\right]\nn
%&&\times\left[\omega_-(k_1)\omega_-(k_2)\omega_-(k_3)\omega_+(k_1+k_2+k_3)
%\right]^{-\frac{1}{2}}
%\label{g}
%\eea
is a symmetric function of its arguments. Its zero momentum limit is
\begin{widetext}
\bea
g(0,0,0)&=&\frac{A_{1-}(0)^3A_{1_+}(0)}{\sqrt{\omega_-^3\omega_+}}
\left[1-\left(\frac{\omega_-^2+H^2-\Delta_1^2}{2H\omega_-}
\right)^2\right] 
\left[1+\frac{\omega_-^2+H^2-\Delta_1^2}{2H\omega_-}
\ \frac{\omega_+^2+H^2-\Delta_1^2}{2H\omega_+}\right]
\equiv\frac{C}{\sqrt{\omega_-^3\omega_+}}.
\label{gsmallk}
\eea
\end{widetext}
where $\omega_\pm=\equiv\omega\pm(0)$.
As a first approximation we neglect all interactions except $V$ and 
calculate the decay rate $M_+\rightarrow 3M_-$ in the Born
approximation.

The differential rate for the decay of a $M_+$ magnon with momentum
$p$ into three $M_-$ magnons with momenta $p_1,p_3,p_3$ is again given
by \r{dGamma}, where
\bea
M&=&\langle0|\prod_{j=1}^3a_-(p_j)\ V\
a_+^\dagger(p)|0\rangle\ \left(\delta(p-\sum_{j=1}^3p_j)\right)^{-1}\nn
&=&\frac{3\lambda v^2}{\pi}g(p_1,p_2,p_3).
\eea
Taking the $M_+$ magnon to be at rest, i.e. setting $p=0$, we obtain
\bea
\Gamma&=&\frac{3\lambda^2 v^4}{\pi}
\int dp_1dp_2|g(p_1,p_2,-p_1-p_2)|^2\nn
&\times& \delta(\omega_+(0)-\omega_-(p_1)-\omega_-(p_2)-\omega_-(p_1+p_2)).
\label{decayrateLG}
\eea
For $H$ slightly larger than $H_{\rm d}$ we again have
\be
\omega_+(0)-3\omega_-(0)\ll \omega_-(0)\ ,
\ee
and the momenta $p_{1,2}$ in \r{decayrateLG} have to be small in order
to satisfy the delta-function. We may use the expansions \r{gsmallk}
for $g$ and 
\bea
&&\omega_-(p)\approx \omega_-(0)+\alpha p^2+{\cal O}(p^4)\ ,\nn
&&\alpha= \frac{v^2}{2\omega_-(0)}\left[1-\frac{2H^2}
{\sqrt{2H^2(\Delta_1^2+\Delta_2^2)+(\frac{\Delta_1^2-\Delta_2^2}{2})^2}}\right].
\quad
\eea
This leads to the following expression for the decay rate in the
regime $H>H_{\rm d}$, $\frac{H}{H_{\rm d}}-1\ll 1$
\bea
\Gamma&\approx&\frac{\sqrt{3}\lambda^2 v^4C^2}{\alpha
  \omega_-(0)^3\omega_+(0)}.  
\label{gammaLG1}
\eea
The result \r{gammaLG1} would suggest that the decay rate switches on
suddenly at a finite value as soon as $H$ becomes larger than $H_{\rm
  d}$. The underlying reason for this jump is the restricted phase
space of a one-dimensional system. 
The ``free boson'' result \r{gammaLG1} for the decay rate is
dramatically different from the result obtained in the framework of
the Majorana fermion theory. This poses the question whether
\r{gammaLG1} is robust if we take into account interactions among the
magnons in the final state. This amounts to resumming the leading
infrared divergences in a perturbative expansion in $\lambda$.
If we assume that in the low-energy limit the $M_-$ degrees of
freedom in the Landau-Ginzburg model can still be mapped onto a Bose
gas with $\delta$-function interactions \cite{ian3}, the result of a
such a resummation can be determined by exploiting the fact
the wave functions of the $M_-$ modes reduce to a free fermion form in
the limit of small momenta \cite{LiebLi}. A three-particle state in
the position representation can be written as  
\bea
|\lambda_1,\lambda_2,\lambda_3\rangle&=&\int
dx_1dx_2dx_3\ \chi(x_1,x_2,x_3)\nn
&&\times\ 
a^\dagger_-(x_1)a^\dagger_-(x_2)a^\dagger_-(x_3)|0\rangle ,
\label{bosegas}
\eea
where the wave-function is given by
\bea
\chi_3(x_1,x_2,x_3)&=&\frac{(-i)^3}{(2\pi)^{3/2}3!}\prod_{j<k=1}^3
\sgn(x_j-x_k)\nn
&&\times \sum_{P\in S_3} \sgn(P) e^{\sum_{j=1}^3i\lambda_{P_j}x_j}.
\eea
Here $P$ denotes a permutation of three elements and $S_3$ the
symmetric group of degree $3$. In momentum space we have
\begin{widetext}
\bea
|\lambda_1,\lambda_2,\lambda_3\rangle&=&\int\frac{dk_1dk_2dk_3}{(2\pi)^3}
\prod_{j=1}^3\frac{2k_j}{k_j^2+\epsilon^2}
\ a^\dagger_-(\lambda_1-k_2-k_3)a^\dagger_-(\lambda_2-k_1+k_3)
a^\dagger_-(\lambda_3+k_1+k_2)|0\rangle .
\label{l123}
\eea
Taking \r{l123} as the final state, the matrix element of the decay
vertex is
\bea
M&=&\frac{\langle\lambda_3, \lambda_2, \lambda_1|V
a_+^\dagger(p)|0\rangle}{\delta(p-\sum_{j=1}^3\lambda_j)}
=\frac{3\lambda v^2}{\pi}\int\prod_{j=1}^3
\frac{dq_j}{2\pi}\frac{2q_j}{q_j^2+\epsilon^2}
\ g(\lambda_1-q_2-q_3,\lambda_2-q_1+q_3,\lambda_3+q_1+q_2)\ .
\label{MLG}
\eea
\end{widetext}
It follows from the definition \r{MLG} that the matrix element is an
antisymmetric function of $\lambda_1,\lambda_2,\lambda_3$. As long as
we are interested in the decay rate for $H$ close to $H_{\rm d}$, we
may expand $M$ for small $\lambda_j$. The leading term antisymmetric
in $\lambda_{1,2,3}$ is then
\bea
C'(\lambda_1-\lambda_2)(\lambda_1-\lambda_3)(\lambda_2-\lambda_3)\ , 
\eea
For small $\lambda_j$ the matrix element is thus equal to
\bea
M&=&\frac{3\lambda C'v^2}{\pi}\prod_{j<k}(\lambda_j-\lambda_k)\ .
\eea
Following through the same steps as before, the decay rate for $H$
close to $H_{\rm d}$ with the $M_+$ magnon initially at rest is found to
be
\bea
\Gamma&\approx&\frac{\sqrt{3}\lambda^2C'^2v^4}{4\alpha^4}
[\omega_+(0)-3\omega_-(0)]^3\propto[H-H_{\rm d}]^3.
\label{gammaLG2}
\eea
Comparing the decay rate \r{gammaLG2} to the result \r{gammaLG1} we
see that interactions in the final state have a dramatic effect:
rather than turning on at a finite value as soon as $H$ exceeds
$H_{\rm d}$, the decay rate \r{gammaLG2} actually vanishes at $H_{\rm
  d}$ and exhibits the same power law behavior for $H>H_{\rm d}$ as
the decay rate in the Majorana fermion model. This may suggest that
the dependence of the decay rate on $H-H_{\rm d}$ is a robust result
that holds for the underlying lattice model as well.

%%%%%%%%%%%%%%%%%%%%%%%%%%%%%%%%%%%%%%%%%%%%%%%%
\section{Vicinity of the Ising Critical Point}
\label{vicinity}
%%%%%%%%%%%%%%%%%%%%%%%%%%%%%%%%%%%%%%%%%%%%%%%%
As we have seen above, both the Majorana fermion model and the
Landau-Ginzburg theory lead to an Ising critical point at some values
$H_c$ of the applied magnetic field. 
\begin{figure}[ht]
\noindent
\epsfxsize=0.45\textwidth
\epsfbox{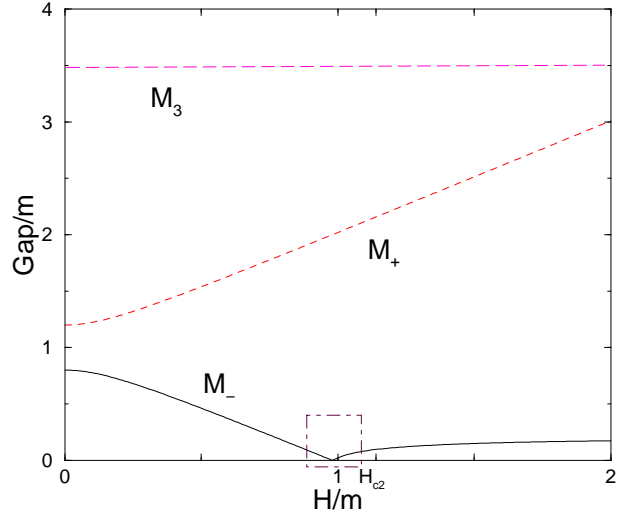}
\caption{Magnon gaps as functions of the applied field for the
Majorana fermion model with $g_a=0$. The gaps in zero field
are chosen as $\Delta_1=0.8m$, $\Delta_2=1.2m$ and
$\Delta_3=3.5m$. The square indicates the window in energies and
fields in which the description in terms of an effective Ising model
is appropriate.
\label{fig:disp}% 
} 
\end{figure}
In the simplest approximations
where interactions are neglected, the actual values for $H_c$ are
rather different, but as we are interested only in robust features this 
numerical difference does not really concern us here. More
importantly both theories predict that the low-energy behaviour in the
vicinity of the critical point is described by an off-critical Ising model
with Hamiltonian
\bea
{\cal H}&=&\frac{iv}{2}\left[L\partial_xL- R\partial_xR\right]
-im_0RL\ .
\label{ising}
\eea
Here the mass parameter that parametrizes the deviation from
criticality is equal to the smallest magnon gap
\be
|m_0|\simeq\omega_-(k=0,H)\ .
\ee
The description by \r{ising} is appropriate at low energies 
$\omega\ll\min\{\omega_+(k=0,H),\omega_3(k=0,H)\}$ and $H$
sufficiently smaller than the critical field $H_{c2}$ \r{hc2} at which
incommensurabilities develop. The Hamiltonian \r{ising} exhibits a
$Z_2$ symmetry
\be
R\longrightarrow -R\ ,\quad
L\longrightarrow -L\ ,
\label{z21}
\ee
under which the Ising order ($\sigma$) and disorder ($\mu$) parameters
transform as
\bea
\underline{H<H_c}:\qquad \sigma&\longrightarrow& -\sigma\ ,\nn
\mu&\longrightarrow& \mu\ ,\nn
\underline{H>H_c}:\qquad \sigma&\longrightarrow& \sigma\ ,\nn
\mu&\longrightarrow& -\mu\ .
\label{z22}
\eea
In order to determine the magnetic low-energy response (in the
vicinity of the antiferromagnetic wave number) we need to
express the staggered magnetizations $n^a$ in terms of operators
related to the Ising model \r{ising}. As was pointed out in
Ref.[\onlinecite{ian2}], the dominant contribution to the staggered
magnetization in the $x$-direction should be the Ising order parameter
field 
\be
n^x={\cal C}_x(H)\ \sigma +\ldots
\label{ops1}
\ee
Here ${\cal C}_x(H)$ is an unknown constant and the dots indicate
contributions from less relevant operators. 
The underlying reason for the identification \r{ops1} is simply that
in the ordered phase of the Ising model (which corresponds to $H>H_c$)
one has a nonzero expectation value $\langle\sigma\rangle\neq 0$,
whereas in the disordered phase (which occurs at $H<H_c$) one has
$\langle\sigma\rangle= 0$. The staggered magnetization in both the
Majorana fermion and LG models exhibits the same kind of behaviour and
it is then natural to expect the identification \r{ops1}.
In Appendix \ref{operatorcontent} we present arguments that suggest that
the staggered magnetization along the $y$-direction has the following
low-energy projection
\bea
n^y={\cal C}_y(H)\ \partial_\tau\sigma +\ldots
\label{ops}
\eea
In what follows we chose a short-distance normalization for the field
$\sigma$ in the theory \r{ising} such that for $\tau^2+x^2/v^2\to 0$
\bea
\langle\sigma(\tau,x)\ \sigma(0,0)\rangle\longrightarrow\frac{1}{[\tau^2+x^2/v^2]^{1/8}}\ .
\label{normalizationsigma}
\eea
We will use the integrability of the theory \r{ising} to determine the
dynamical structure factor $S^{\rm xx}(\omega, \frac{\pi}{a_0}+q)$ at
low energies, where the staggered component of $S^{x}$ is given by
\r{ops1}. The $\rm yy$ component then follows from \r{ops} 
\bea
S^{\rm yy}(\omega,\frac{\pi}{a_0}+q)\propto
\frac{\omega^2}{m^2} S^{\rm xx}(\omega,\frac{\pi}{a_0}+q)\ .
\eea

%%%%%%%%%%%%%%%%%%%%%%%%%%%%%%%%%%%%%%%%
\subsubsection{$H<H_c$: Low-Field Phase}
%%%%%%%%%%%%%%%%%%%%%%%%%%%%%%%%%%%%%%%%
By the $Z_2$ symmetry \r{z21},\r{z22} only intermediate states with an
odd number of magnons contribute to the two-point correlation
functions of the Ising order parameter field $\sigma$ and hence the
staggered structure factor $S^{\rm xx}(\omega,\frac{\pi}{a_0}+q))$ at
low energies. The leading contributions are \cite{barouch,yurov,cardy,fab}
\bea
&&S^{\rm xx}(\omega,\frac{\pi}{a_0}+q)=
\frac{vA}{\sqrt{m_0^2+v^2q^2}}
\delta\left(\omega-\sqrt{m_0^2+v^2q^2}\right)\nn
&&+\frac{2vA}{3\pi^2m_0^2}\int_0^{z_0}dz\frac{\left(\tanh(z)\tanh(\frac{y+z}{2})
\tanh(\frac{y-z}{2})\right)^2}
{\sqrt{[x^2-1-4\cosh^2z]^2-16\cosh^2z}}\nn
&& +\ \text{\rm contributions from 5,7,\ldots magnons},
\label{DSFlow}
\eea
where 
\bea
A&=&{\cal C}_x^2(H)2^\frac{1}{6}e^{-\frac{1}{4}}{\cal A}^3\
m_0^\frac{1}{4}\ ,
\label{A}\\
x^2&=&\frac{\omega^2-v^2q^2}{m_0^2}\ ,\nn
z_0&=&{\rm arccosh}(\frac{x-1}{2})\ ,\nn
y&=&{\rm arccosh}\left(\frac{x^2-1-4\cosh^2z}{4\cosh z}\right)\ .
\eea
Here 
\bea
A=1.28242712910062...
\label{glaisher}
\eea
is Glaisher's constant.
The three-magnon contribution is always very small. 
In the frequency interval $[0,30 m_0]$ roughly 100 times more spectral
weight sits in the single-magnon contribution than in the
three-particle one. Hence the magnetic response below energies of the
order of tens of the magnon gap $m_0$ is dominated by the coherent
magnon contribution. However, if $H$ becomes very close to $H_c$ the
magnon gaps $m_0$ tends to zero. If we are interested in the
magnetic response at a low (compared to the gap of the second coherent
magnon mode) but fixed energy we have to take the contributions
of intermediate states with 5,7,9,... magnons into account in order to
get an accurate result for $S^{\rm xx}(\omega,\frac{\pi}{a_0}+q)$.

%%%%%%%%%%%%%%%%%%%%%%%%%%%%%%%%%%%%%%%%
\subsubsection{$H=H_c$: At Criticality}
%%%%%%%%%%%%%%%%%%%%%%%%%%%%%%%%%%%%%%%%
At criticality the structure factor exhibits a power-law
behaviour \cite{sachdev}
\bea
S^{\rm xx}(\omega,\frac{\pi}{a_0}+q)&=&{\rm Im}\left\lbrace
{\cal B}\left[v^2q^2-(\omega+i\varepsilon)^2\right]^{-\frac{7}{8}}\right\rbrace ,
\eea
where
\be
{\cal B}=2v\ {\cal C}^2_x(H)
2^\frac{3}{4}\frac{\Gamma\left(\frac{7}{8}\right)}{\Gamma\left(\frac{1}{8}\right)}\ .
\ee
%%%%%%%%%%%%%%%%%%%%%%%%%%%%%%%%%%%%%%%%%%
\subsubsection{$H>H_c$: High-field Phase}
%%%%%%%%%%%%%%%%%%%%%%%%%%%%%%%%%%%%%%%%%%
In this phase there is a nonzero staggered magnetization and
correspondingly a nonzero expectation value for the Ising order
parameter
\be
\langle \sigma \rangle\neq 0,
\ee
which results in a Bragg peak for momentum transfer $\frac{\pi}{a_0}$
along the chain direction. By virtue of the $Z_2$ symmetry \r{z22}
only intermediate states with an even number of magnons contribute to
$S^{\rm xx}(\omega,\frac{\pi}{a_0}+q)$. The leading contribution to the
inelastic neutron scattering cross section comes from intermediate
states involving {\sl two} magnons \cite{barouch,yurov,cardy}, which
leads to the following result
\bea
&&S^{\rm xx}(\omega,\frac{\pi}{a_0}+q)=\frac{vA}{\pi}\frac{
\sqrt{s^2-4m_0^2}}{s^3}\theta(s-2m_0)\nn
&& +\ \text{\rm contributions from 4,6,\ldots magnons}\ .
\label{DSFhigh}
\eea
Here $A$ is given by \r{A} and $s^2=\omega^2-v^2q^2$.

%%%%%%%%%%%%%%%%%%%%%%%%%%%%%%%%%%%%%%%%%%%%%%%%%%%
\section{The High-Field Phase for weak anisotropy}
\label{highfield}
%%%%%%%%%%%%%%%%%%%%%%%%%%%%%%%%%%%%%%%%%%%%%%%%%%%
In general it is difficult to determine the effects of
magnon-interactions in quantitative detail. An exception is the case
of a small anisotropy of the zero-field gaps
\be
\Delta\ll m\ ,
\ee
where $2m=\Delta_1+\Delta_2$ and $\Delta=\Delta_2-\Delta_1$.
As we will show, if the magnetic field is sufficiently larger than the
critical field $H_c$ and 
\be
\Delta\ll H-H_c\alt J\ ,
\ee
it is possible to determine the interaction effects on the magnetic
response at low energies in some detail. More precisely, we consider
magnetic fields sufficiently larger than the field $H_{c2}$ (see
e.g. \r{hc2}), at which incommensurabilities begin to develop.
The scale defining the low-energy region is the difference
$\Delta$. As is shown in Appendix \ref{sineG}, the low-energy
effective Hamiltonian is given by a sine-Gordon model (SGM)\cite{ian2}
\bea
{\cal H}&=&\frac{\tilde{v}}{16\pi}\left[(\partial_x\Theta)^2+
(\partial_x\Phi)^2\right]-2\mu\cos\beta\Theta\ .
\label{SGM}
\eea
Here $\tilde{v}$ is the spin velocity, $\mu\propto \Delta$ and $\beta$
is a function of the applied magnetic field $H$. The high-energy
cutoff of the theory \r{SGM} is $H-H_c$. At low energies the dominant
Fourier component of the transverse spin operators is at $q=\frac{\pi}{a_0}$ 
\bea
S_j^\pm&\longrightarrow& (-1)^j A\exp\left(\pm
\frac{i\beta}{2}\Theta\right)\ .
\label{spins}
\eea
The identification \r{spins} is in accordance with the fact that in
the high-field phase there is N\'eel order along the x-direction
\be
\bigl\langle(-1)^nS^x_n\bigr\rangle\propto
\bigl\langle\cos\bigl(\frac{\beta}{2}\Theta\bigr)\bigr\rangle\neq 0\ .
\ee
We choose a short-distance normalization such that
for $|x-y|\to 0$
\be
e^{i\alpha\Theta(x)}e^{i\gamma\Theta(y)}\longrightarrow
|x-y|^{4\alpha\gamma}e^{i\alpha\Theta(x)+i\gamma\Theta(y)}.
\label{VON}
\ee
The amplitude $A$ in \r{spins} is nonuniversal and not known in
general. However,  very close to $H_c$
\be
\Delta\ll H-H_c\ll H_c,
\ee
it is given by (see Appendix \ref{app:corramp})
\be
A=A'\left[\frac{a_0^4m}{v^2}(H-H_c)\right]^\frac{1}{8},
\label{corramp}
\ee
where $A'$ is a field independent numerical constant. The dependence
of \r{corramp} on $H-H_c$ is a {\sl universal feature of the C-IC
transition}.

We note that the sign of the $\cos$-term in \r{SGM} is quite
important. Flipping the sign corresponds to a shift
$\Theta\longrightarrow\Theta+\pi/\beta$, which essentially leads to an
exchange of the $x$ and $y$ component of the spin operators in \r{spins}.

The value of the parameter $\beta$ is of crucial importance. For
the isotropic case ($\Delta_2=\Delta_1$) it has recently been determined
\cite{KF} in the framework of the nonlinear sigma model description of
the spin-S Heisenberg chain. In the isotropic case the high-field
phase at $H>H_c$ is a Luttinger liquid and $\beta$ is related to the
Luttinger liquid parameter. It was found that \cite{KF}
\be
\beta=\frac{1}{\sqrt{2}S_R(\theta_F)}\ ,
\label{beta}
\ee
where $S_R(\theta)$ fulfils the integral equation 
\bea
S_R(\theta)&=&1+\int_{-\theta_F}^{\theta_F}d\theta'\ S_R(\theta')\ 
\frac{1}{\pi^2+(\theta-\theta')^2}\ .
\eea
Here $\theta_F$ is determined as a function of the magnetic field $H$
by
\bea
\epsilon(\theta)&=&m\cosh(\theta)-H+\int_{-\theta_F}^{\theta_F}d\theta' 
\epsilon(\theta')\frac{1}{\pi^2+(\theta-\theta')^2},\nn
\epsilon(\theta_F)&=&0.
\eea
\begin{figure}[ht]
\noindent
\epsfxsize=0.42\textwidth
\epsfbox{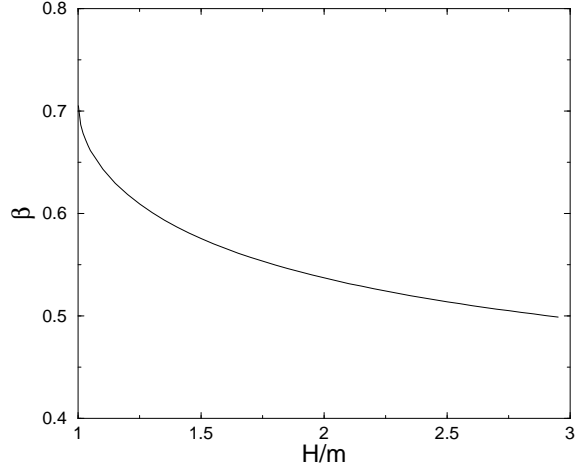}
\caption{Parameter $\beta$ as function of the applied magnetic field.
\label{fig:beta}% 
} 
\end{figure}
Similarly one may determine the spin velocity
\bea
\tilde{v}=\frac{v}{2\pi\rho(\theta)}
\frac{\partial\epsilon(\theta)}{\partial\theta}\biggr|_{\theta=\theta_F},
\eea
where $\rho(\theta)$ fulfils the integral equation
\be
\rho(\theta)=\frac{m}{2\pi}\cosh(\theta)+
\int_{-\theta_F}^{\theta_F}d\theta'  
\rho(\theta')\frac{1}{\pi^2+(\theta-\theta')^2}.
\ee

The parameter $\beta$ as well as the velocity $\tilde{v}$ entering the
sine-Gordon Hamiltonian \r{SGM} may be estimated with a good degree of
accuracy from their respective values in the isotropic case as long as 
$\frac{\Delta_2-\Delta_1}{\Delta_2}\ll 1$. On the other hand, the
agreement between the Luttinger liquid parameter calculated in the
nonlinear sigma model and the one of the isotropic spin-1 Heisenberg
chain in a magnetic field, which is known approximately from DMRG
computations \cite{campos}, was shown to be fairly good in
Ref.[\onlinecite{KF}]. Hence we may determine $\beta$ with a
reasonable degree of accuracy from \r{beta} as long
$\frac{\Delta_2-\Delta_1}{\Delta_2}\ll 1$. It then follows that
$\beta<\frac{1}{\sqrt{2}}$ and hence the SGM \r{SGM} is in the
attractive regime.
%%%%%%%%%%%%%%%%%%%%%%%%%%%%%%%%%%%%%%%%
\subsection{Spectrum of the SGM}
%%%%%%%%%%%%%%%%%%%%%%%%%%%%%%%%%%%%%%%%
The SGM \r{SGM} is integrable and many exact results are available.
The spectrum of the SGM depends on the value of the coupling
constant~$\beta$. In the so-called repulsive regime, $1/\sqrt{2}<\beta<1$,
there are only two elementary excitations, called soliton
and antisoliton. These have a massive relativistic dispersion,
\begin{equation}
E(P)=\sqrt{M^2+\tilde{v}^2 P^2} \; ,
\label{holondispersion}
\end{equation}
where $M$ is the gap.

In the regime $0<\beta <1/\sqrt{2}$ relevant to our discussion, 
soliton and antisoliton attract and can form bound states, known as
``breathers''. There are  
\begin{equation}
{\cal N}= \left[ \frac{1-\beta^2}{\beta^2}\right]
\label{Nex}
\end{equation}
different types of breathers, where $[ x ]$ in~(\ref{Nex}) denotes
the integer part of~$x$. The breather gaps are given by
\begin{equation}
M_n=2M \sin(n\pi\xi/2)\ ,\quad n=1,\ldots ,{\cal N}\; ,
\label{exgap}
\end{equation}
where 
\begin{equation}
\xi=\frac{\beta^2}{1-\beta^2} \; .
\label{xi}
\end{equation}
The number of breathers is a function of the applied magnetic field
$H$. There always is at least one breather. A second breather appears
above a field $H_0$, which is determined by the requirement
\be
\beta=\frac{1}{\sqrt{3}}.
\ee
This requirement is fulfilled for
\be
H>H_0\approx 1.5M\ .
\ee

%%%%%%%%%%%%%%%%%%%%%%%%%%%%%%%%%%%%%%%%
\subsection{Dynamical Structure Factor}
%%%%%%%%%%%%%%%%%%%%%%%%%%%%%%%%%%%%%%%%

A basis of eigenstates of the SGM is given by scattering states of
solitons, antisolitons and breathers. In order to distinguish these we
introduce labels $B_1,B_2,\ldots B_{\cal N},s,\bar{s}$. As usual
for particles with relativistic dispersion, it is useful to introduce
a rapidity variable $\theta$ to parameterize energy and momentum
\begin{eqnarray}
&&E_{s}(\theta)=M\cosh\theta\; , \; P_{s}(\theta)=(M/\tilde{v})
\sinh\theta\; ,\\[3pt]
&&E_{\bar{s}}(\theta)=M\cosh\theta\; , \; 
P_{\bar{s}}(\theta)=(M/\tilde{v})\sinh\theta \; ,\\[3pt]
&&E_{B_n}(\theta)=M_n\cosh\theta\; , \; 
P_{B_n}(\theta)=(M_n/\tilde{v})\sinh\theta\; .\qquad
\end{eqnarray}
Two-point functions are expressed in terms of a basis of scattering
states of solitons, antisolitons and breathers as summarized in
Appendix \ref{app:spectral}, see Eqn \r{2pt}.
After carrying out the double Fourier transform we arrive at the
following representation for the imaginary part of the retarded
two-point function of the operator ${\cal O}$ for $\omega>0$
\begin{widetext}
\begin{eqnarray}
\label{expansion1}
S^{\cal O}(\omega,q)= 
\sum_{n=1}^\infty\sum_{\epsilon_i}\int
\frac{d\theta_1\ldots d\theta_n}{(2\pi)^{n-1}n!}
|f^{\cal O}_{\epsilon_1\ldots\epsilon_n}(\theta_1\ldots\theta_n)|^2 
\delta(q-\sum_jM_{\epsilon_j}\sinh\theta_j/\tilde{v})\
\delta(\omega-\sum_jM_{\epsilon_j}\cosh\theta_j)
\mbox{}
\end{eqnarray}
\end{widetext}
The form factors of the operators $\exp(\pm i\beta\Phi/2)$ in the
sine-Gordon model were determined in Ref.~[\onlinecite{smirnov,lukyanov}].
Using these results we can determine the first few terms of the
expansion~(\ref{expansion1}) for the transverse spin operators. We
have  
\bea
S^{\rm xx}(\omega,\frac{\pi}{a_0}+q)&=&C\Biggl\{
\pi f_2\delta(s^2-M_2^2)\Theta(H-H_0)\nn
&&+
{\rm Re}\frac{|F^{\rm
cos}(\theta_0)|^2}{s\sqrt{s^2-4M^2}}+\ldots\Biggl\},\nn
S^{\rm yy}(\omega,\frac{\pi}{a_0}+q)&=&C\Biggl\{
\pi f_1\delta(s^2-M_1^2)\nn
&&+{\rm Re}\frac{|F^{\rm
sin}(\theta_0)|^2}{s\sqrt{s^2-4M^2}}+\ldots\Biggr\}.
\label{DSF_SG}
\eea
Here $C$ is an overall (dimensionful) constant. The terms proportional
to $F^{\rm sin}$ and $F^{\rm cos}$ represent the contributions by
intermediate states involving one soliton and one antisoliton and
\be
\theta_0=2{\rm arccosh}(s/2M).
\ee
The $\delta$-function contributions are due to the
breather bound states. The soliton-antisoliton form factors are given
by \cite{smirnov,lukyanov} 
\bea
|F^{\rm sin}(\theta)|^2&=&\Bigl|\langle
0|\sin\left(\frac{\beta}{2}\Phi(0)\right)|\theta_2\theta_1\rangle_{+-}\Bigr|^2\nn
&=&\Biggl|\frac{g(\theta)}{\xi\cosh\left(
\frac{\theta+i\pi}{2\xi}\right)}\Biggr|^2\ ,\nn
|F^{\rm cos}(\theta)|^2&=&\Bigl\langle
0|\cos\left(\frac{\beta}{2}\Phi(0)\right)|\theta_2\theta_1\rangle_{+-}\Bigr|^2\nn
&=&\Biggl|\frac{g(\theta)}{\xi\sinh\left(
\frac{\theta+i\pi}{2\xi}\right)}\Biggr|^2,
\eea
where $\theta=\theta_2-\theta_1$ and
\bea
&&g(\theta)=i\sinh{\theta/2}\nn
&&\times \exp \left ( \int_0^\infty\frac{dt}{t}
\frac{\sinh^2(t[1-i\theta/\pi])\sinh(t[\xi-1])}{\sinh 2t\ \sinh\xi t\ \cosh t}\right).\nn
\eea
The absolute values squared of the breather form factors are
\cite{smirnov,lukyanov} 
\bea
f_1&=&2\sin\left(\frac{\pi\xi}{2}\right)
\exp\left(-2\int_0^{\pi\xi}\frac{dt}{2\pi}\frac{t}{\sin t}\right),\nn
f_2&=&\frac{2|g(-i\pi[1-2\xi])|^2}{\cot(\pi\xi)\cot(\frac{\pi\xi}{2})^2}\ .
\eea
An important result is that the first bound state $B_1$ is visible
only in $S^{yy}$ and does not couple to $S^{xx}$. We note that there
are additional contributions in the spectral representations
\r{DSF_SG} at higher energies. For example, there is a two-breather
$B_1B_1$ contribution to $S^{xx}$ at energies above $2M_1$.

\begin{figure}[ht]
\noindent
\epsfxsize=0.45\textwidth
\epsfbox{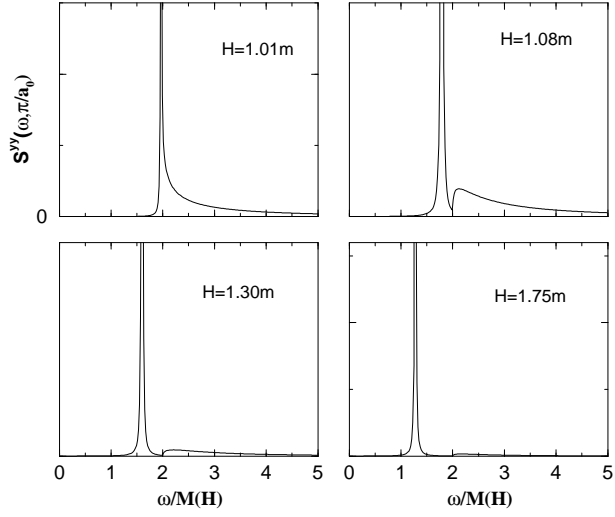}
\caption{$S^{\rm yy}(\omega,\frac{\pi}{a_0})$ as a function of
$\omega/M$ for several values of the applied field $H$.
\label{fig:syy}% 
} 
\end{figure}

\begin{figure}[ht]
\noindent
\epsfxsize=0.45\textwidth
\epsfbox{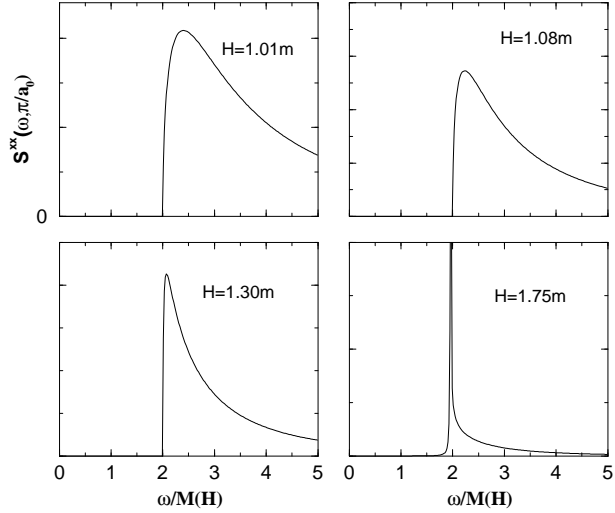}
\caption{$S^{\rm xx}(\omega,\frac{\pi}{a_0})$ as a function of
$\omega/M$ for several values of the applied field $H$.
\label{fig:sxx}% 
} 
\end{figure}

We plot $S^{\alpha\alpha}(\omega,\frac{\pi}{a_0})$ as functions of
$\omega/M$ for several values of the applied magnetic field in Figs
\ref{fig:syy}-\ref{fig:sxx}. In order to give a visual impression of
their spectral weights we have broadened the $\delta$-functions
corresponding to the breathers by convolution with a Gaussian.

We first discuss the evolution of $S^{\rm yy}(\omega,\frac{\pi}{a_0})$
shown in Fig. \ref{fig:syy}: as $H$ moves away from $H_c=m$ the
breather $B_1$ splits off from the soliton-antisoliton continuum and
very quickly takes over most of the spectral weight. Except for a
narrow window (in magnetic field) above $H_c$ the $\rm yy$-component
of the dynamical structure factor is dominated by a {\sl coherent}
single-particle peak.

The evolution of $S^{\rm xx}(\omega,\frac{\pi}{a_0})$ is very
different as is shown in Fig. \ref{fig:sxx}: as $H$ moves away from
$H_c=m$ the incoherent soliton-antisoliton continuum slowly narrows
(on the scale of the field-dependent soliton gap)
until it eventually begets the second, heavy breather $B_2$ at
$H=H_0\approx 1.5m$. Over a large interval of magnetic fields 
$S^{\rm xx}(\omega,\frac{\pi}{a_0})$ is dominated
by the {\sl incoherent} soliton-antisoliton continuum.

%%%%%%%%%%%%%%%%%%%%%%%%%%%%%%%%
\subsubsection{Spectral Weights}
\begin{figure}[ht]
\noindent
\epsfxsize=0.45\textwidth
\epsfbox{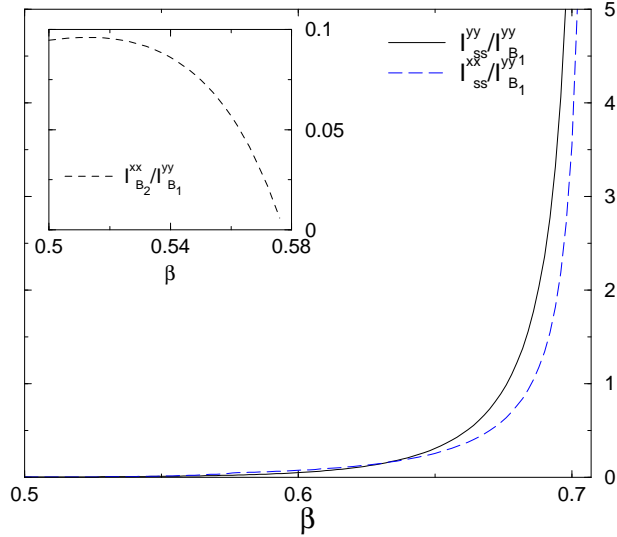}
\caption{Spectral weight ratios $I^{\rm yy}_{ss}/I^{\rm yy}_{B_1}$, 
$I^{\rm xx}_{ss}/I^{\rm yy}_{B_1}$ and $I^{\rm xx}_{B_2}/I^{\rm
    yy}_{B_1}$ as functions of the parameter $\beta$.
\label{fig:SW}% 
} 
\end{figure}

%%%%%%%%%%%%%%%%%%%%%%%%%%%%%%%%
In order to compare the spectral weights located in the coherent
breather peaks to the spectral weight associated with the
soliton-antisoliton continua it is useful to define quantities 
\bea
I^{\rm xx}&=&\frac{M^2}{C}\int_0^{25}dx\ S^{\rm xx}(xM,\frac{\pi}{a_0})
\equiv I^{\rm xx}_{B_2}+I^{\rm xx}_{ss},\nn
I^{\rm yy}&=&\frac{M^2}{C}\int_0^{25}dx\ S^{\rm yy}(xM,\frac{\pi}{a_0})
\equiv I^{\rm yy}_{B_1}+I^{\rm yy}_{ss}.\nn
\eea
For example, $CI^{\rm yy}/M^2$ is the spectral weight of the
yy-component of the dynamical structure factor at the
antiferromagnetic wave number integrated over the frequency interval
$[0, 25M]$. It has contributions $I_{B_1}^{\rm yy}$
from the coherent breather peak and $I_{ss}^{\rm yy}$ from the soliton
antisoliton continuum (there are also contributions to due $B_1B_2$
two-breather states {\sl etc}, but their contributions are
subleading). It is important to note that the soliton gap $M$ and the
overall factor $C$ depend on the applied magnetic field. These
dependencies drop out once we consider spectral weight ratios such as 
\be
\frac{I^{\rm yy}_{ss}}{I^{\rm yy}_{B_1}}\ ,\quad
\frac{I^{\rm xx}_{ss}}{I^{\rm yy}_{B_1}}\ ,\quad
\frac{I^{\rm xx}_{B_2}}{I^{\rm yy}_{B_1}}\ .
\ee
These ratios are plotted as functions of $\beta$ in Fig.\ref{fig:SW}.
We see that for small $\beta$ (that is at $H\gg H_c$) most of the
spectral weight is situated in the coherent peak associated with the
first breather $B_1$. Very close to the transition the second breather
does not exist and most of the spectral weight sits in the
soliton-antisoliton continua. The crossover between these two regimes
occurs around $\beta\approx 0.675$, which according to
Fig. \ref{fig:beta} corresponds to
\be
\frac{H}{H_c}\approx 1.025 \ .
\ee
The lesson is that interactions make the summed dynamical structure
factor 
\be
S^{\rm xx}(\omega,\frac{\pi}{a_0}+q)+S^{\rm
  yy}(\omega,\frac{\pi}{a_0}+q)
\ee
look coherent expect for fields very close to $H_c$.
On the other hand, the polarized structure factor
$S^{\rm xx}(\omega,\frac{\pi}{a_0}+q)$ looks incoherent!
It would be very interesting to attempt to disentangle the components
of the dynamical structure factor in inelastic neutron
scattering experiments and in this way observe this incoherent
scattering continuum.

%%%%%%%%%%%%%%%%%%%%%%%%%%%%%%%%%%%%%%%%%%%%%%%%%
\subsubsection{Polarizations in the LG model}
%%%%%%%%%%%%%%%%%%%%%%%%%%%%%%%%%%%%%%%%%%%%%%%%%
How do these results fit into the general picture of the LG model?
In the latter one expands to quadratic order in the fields $\varphi_1$
and $\varphi_2$ around the minimum of the effective potential at
$\vec{\varphi}_{\rm   vac}=(m_0,0,0)$   
\be
m_0^2=\frac{H^2-\Delta_1^2}{4v\lambda}\ .
\ee
The effective Lagrangian for the fields $\varphi_{1,2}$ becomes
\bea
{\cal L}&=&\sum_{a=1}^2
\frac{1}{2v}\left(\frac{\partial\varphi_a}{\partial t}\right)^2
-\frac{v}{2}\left(\frac{\partial\varphi_a}{\partial x}\right)^2
-\frac{H_c}{v}\ 
\epsilon_{abc}\frac{\partial\varphi_a}{\partial t}\ \varphi_b\nn
&&-\frac{H^2-\Delta_1^2}{v}\varphi_1^2-\frac{\Delta_2^2-\Delta_1^2}{2v}\varphi_2^2\ .
\eea
This is the same as \r{LG} for the fields $\varphi_{1,2}$ if we make
the replacement (in \r{LG})
\bea
\Delta_1^2&\longrightarrow& 3H^2-2\Delta_1^2\ ,\nn
\Delta_2^2&\longrightarrow& \Delta_2^2+H^2-\Delta_1^2\ .
\eea
This implies for the polarizations in the limit $H\gg \Delta_2$
\bea
\frac{A_{2-}(0)}{A_{1-}(0)}&\longrightarrow
-\left[\frac{3H^2}{\Delta_2^2-\Delta_1^2}
\right]^\frac{1}{2}\ .
\eea
In other words
\bea
|A_{2-}|\gg |A_{1-}|\ ,
\eea
and as a result the coherent low-energy mode is dominantly polarized
along the y-direction! This agrees nicely with the sine-Gordon
calculation, where the dominant feature, the first breather, appears
in $S^{\rm yy}$.

%%%%%%%%%%%%%%%%%%%%%%%%%%%%%%%%%%%%%%
\section{Interchain Coupling}
\label{interchain}
%%%%%%%%%%%%%%%%%%%%%%%%%%%%%%%
So far we have considered a purely one-dimensional situation
corresponding to an ensemble of uncoupled spin-1 chains. As long as
the magnon gap is large, a weak coupling between the chains may be
neglected in a first approximation. On the other hand, the interchain
exchange is expected to lead to significant qualitative changes in the
magnetic response close to the critical point where the magnon gap
becomes very small \cite{Taka1}. In order to assess the effects of a weak
interchain coupling for $H\approx H_c$ we consider a Landau-Ginzburg
model of the form
\bea
{\cal L}=\sum_n{\cal L}_n+{\cal L}_{\rm int}\ ,
\label{Lcoupled}
\eea
where
\bea
{\cal L}_n&=&\frac{1}{2v}\left(
\frac{\partial\vec{\varphi}_n}{\partial t}+\vec{H}\times\vec{\varphi}_n\right)^2
-\frac{v}{2}\left(
\frac{\partial\vec{\varphi}_n}{\partial x}\right)^2\nn
&&-\sum_{a=1}^3\frac{\Delta_a^2}{2v}\varphi^2_{n,a}-\lambda|\vec{\varphi}_n|^4,\nn
{\cal L}_{\rm  int}&=&\frac{J_\perp}{a_0}
\sum_{\langle j,k\rangle}\vec{\varphi}_j\cdot\vec{\varphi}_{k}\ .
\label{LGn}
\eea
Here the sum $\langle jk\rangle$ is over links between neighbouring
sites on different chains and we have dropped quartic terms in ${\cal
L}_{\rm int}$ that arise from the interaction of the smooth components
of the spin operators. 

As we have seen in section \ref{vicinity}, close to the the Ising
critical point the low-energy degrees of freedom are described by
off-critical Ising models. Hence at low energies we have
\bea
{\cal L}_n&\approx&R_n\partial_- R_n+L_n\partial_+ L_n -i m_0\ R_nL_n\ .
\label{coupled}
\eea
The leading low-energy projection of the interchain exchange follows
from \r{ops1}, \r{ops}
\be
{\cal L}_{\rm int}\approx\frac{J_\perp}{a_0} C^2\left[\frac{a_0}{v}\right]^{1/4}
\sum_{\langle j,k\rangle} \sigma_j\ \sigma_{k}\ ,
\label{coupled2}
\ee
where $C$ is a dimensionless constant. The ``quasi-1D Ising model''
\r{coupled}, \r{coupled2} has recently been studied in
Ref. [\onlinecite{CT}] and we may follow some of this analysis here.

%%%%%%%%%%%%%%%%%%%%%%%%%%%%%%%%%%%%%%%
\subsection{Mean-Field Approximation}
%%%%%%%%%%%%%%%%%%%%%%%%%%%%%%%%%%%%%%%

As a first step, we analyze the model \r{coupled}, \r{coupled2} by
means of a self-consistent mean-field approximation
\cite{Pincus,Taka2}. We assume the existence of a nonzero expectation
value 
\bea
\langle \sigma \rangle\ \neq 0\ ,
\label{stagmag}
\eea
which corresponds to N\'eel order along the x-direction. The
long-range order can be induced by the magnetic field, the
interchain coupling or by both. In the presence on a nonzero
expectation value \r{stagmag} we may decouple the interaction term in
\r{coupled2} and arrive at the following mean-field Lagrangian density 
\bea
{\cal L}_{\rm MF}&=&
R\partial_- R+L\partial_+ L -i m_0\ RL +
%\left[\frac{v}{a_0}\right]^\frac{15}{8}\frac{h}{v}\
\frac{h}{v}\
\sigma\ .\nn
\label{Ising}
\eea
Here ``magnetic field'' $h$ has dimensions of $s^{-\frac{15}{8}}$ by
virtue of the normalization \r{normalizationsigma} and is subject to the
self-consistency condition
\bea
h={\cal Z}C^2v\frac{J_\perp}{a_0} 
%\left[\frac{a_0}{v}\right]^{\frac{17}{8}}\
\left[\frac{a_0}{v}\right]^{\frac{1}{4}}\
\langle\sigma\rangle\ ,
\label{SCh}
\eea
where $\cal Z$ is the number of neighbouring chains.
The mean-field theory is purely one-dimensional and describes
an off-critical Ising model in an effective magnetic field induced by
the neighbouring chains. The model \r{Ising} has been studied by
several authors \cite{McCoyWu1,McCoyWu2,Mussardo} and is known to
exhibit very interesting physical behaviour as $m_0$ and $h$ are
varied. In order to discuss the effects of $h$ and $m_0$ it is
convenient to consider the Euclidean two-point function of Ising order
parameters
\be
\chi^E_{\sigma\sigma}(\bar\omega,q)=\int_{-\infty}^\infty dx\ d\tau\
e^{i\bar\omega\tau-iqx}\ \langle\sigma(\tau,x)\sigma(0,0)\rangle\ .
\label{chiE}
\ee
We note that $\chi^E_{\sigma\sigma}$ is related to the xx-component of
the staggered susceptibility by analytic continuation to real frequencies.
The Lagrangian \r{Ising} defines a one-parameter family of field
theories labelled by the dimensionless quantity \cite{Mussardo}
\be
\chi=m_0h^{-\frac{8}{15}}\ .
\ee
In the two special cases $\chi=0$ and $|\chi|=\infty$ the
model\r{Ising} is integrable and the susceptibility \r{chiE} can be
determined to a very high accuracy by means of the formfactor
bootstrap approach. In what follows we first review
known quantitative results for the cases $|\chi|\to\infty$ and
$\chi\to 0$ and then summarize the qualitative behaviour for general
values of $\chi$.

%%%%%%%%%%%%%%%%%%%%%%%%%%%%%%%%%%%%%%%%%%%%%%%%%%%
\subsubsection{The Limit $h\to 0$: McCoy-Wu Scenario}
%%%%%%%%%%%%%%%%%%%%%%%%%%%%%%%%%%%%%%%%%%%%%%%%%%%
The regime $h\to 0$ was studied in Refs \cite{McCoyWu1,McCoyWu2} by
means of a perturbative expansion in $h$. In the absence of a field
($h=0$) the dynamical structure factor has been given in section
\ref{vicinity}. For $m_0>0$ the Ising model is in its disordered phase
and the spin-spin correlation functions are dominated by a
single-particle pole and the next-lowest excited states occur in the
form of a three-particle scattering continuum. The dynamical structure
factor is proportional to \r{DSFlow}. Introducing a small magnetic
field leads to a small shift in the position of the single-particle
pole. Furthermore a two-particle scattering continuum of excited
states emerges.

For $m_0<0$ the Ising model is in its ordered
phase. This means that there is a nonzero value for
the staggered magnetization
\bea
\langle\sigma\rangle_0^2&=&2^{1/6}e^{-1/4}A^3 m_0^{1/4}\ ,
\eea
where $A$ denotes Glaisher's constant \r{glaisher}.
The structure factor in the ordered phase is given by
\r{DSFhigh}: the structure factor is incoherent and there is a
two-particle branch cut starting at $\omega=2m_0$.

It is convenient to define a dimensionless magnetic field $\tilde{h}$ by
\bea
\tilde{h}&=&\frac{\langle\sigma\rangle_0}{m_0^2}h.
%\left[\frac{v}{m_0a_0}\right]^{\frac{15}{8}}
%A^{3/2}2^{1/12}e^{-1/8}\ h\ .
\eea
After a resummation of a perturbative expansion in $\tilde{h}$ McCoy
and Wu established that the spin-spin correlation function has the
following large-distance behaviour \cite{McCoyWu1}
\bea
\langle\sigma(\tau,x)\ \sigma(0,0)\rangle&\approx&
\langle\sigma\rangle_0^2\ \frac{\exp(-2m_0r)}{2\sqrt{\pi m_0r}}\ \tilde{h}\nn
&\times& \sum_l\exp(-m_0r(\lambda_l\tilde{h})^{2/3})\ ,
\label{sigmasigma}
\eea
where $r^2=\tau^2+x^2/v^2$ and $\lambda_l$ are the positive solutions
to the equation
\bea
J_\frac{1}{3}(\lambda_l/3)+J_{-\frac{1}{3}}(\lambda_l/3)=0\ .
\eea
%The first few solutions $\lambda_l$ are
%\bea
%\lambda_1&\approx& 7.15034...\nn
%\lambda_2&\approx& 16.5306...\nn
%\lambda_3&\approx& 25.9421...\nn
%\lambda_4&\approx& 35.3605...\ .
%\eea
The interesting point is that the Fourier transform of \r{sigmasigma}
no longer has a branch cut! There are single-particle poles at
\bea
\bar\omega^2+v^2q^2=-[2+(\tilde{h}\lambda_l)^{2/3}]^2m_0^2\ .
\eea
In other words, the two-particle branch cut has disintegrated into a
series of single-particle poles. The residues of these poles are 
proportional to
\be
\tilde{h}\ [2+(\tilde{h}\lambda_l)^{2/3}]^{-1}.
\ee
Hence the lightest particle carries more spectral weight than the
heavier ones. This is quite different from the result for $h=0$ where
the structure factor vanishes as the threshold is approached from
above. 

%%%%%%%%%%%%%%%%%%%%%%%%%%%%%%%%%%%%%%%%%%%%%%%%%%%%%%%%
\subsubsection{The Limit $m_0\to 0$: Magnetic Deformation}
%%%%%%%%%%%%%%%%%%%%%%%%%%%%%%%%%%%%%%%%%%%%%%%%%%%%%%%%
In the limit $m_0\to 0$ the model \r{Ising} is integrable
\cite{Zamo89}. The spectrum consists of eight types of massive
self-conjugate particles. Three of them have masses below the lowest
two-particle threshold. The two-point function of the Ising order
operator $\sigma$ was calculated by the form factor bootstrap approach
in Ref.[\onlinecite{Delfino}]. The dominant contribution to the two
point function of Ising order parameter fields is due to the lightest
particles. The dynamical susceptibility is approximately
\be
\chi_{\sigma\sigma}(\omega,q)\approx 
\left[\frac{4m_1^2}{15\pi
    h}\right]^2\sum_{j=1}^3\frac{2vZ_j}{\omega^2-v^2q^2-m_j^2}\ ,
\ee
where the particle masses are \cite{Zamo89,Fateev}
\bea
m_1&\approx& 4.40490858\ h^\frac{8}{15}\ ,\nn
m_2&\approx& 1.618\ m_1\ ,\quad
m_3\approx 1.989\ m_1.
\eea
and \cite{Delfino,CT}
\bea
Z_1\approx 0.247159\ ,\
Z_2\approx 0.069017\ ,\
Z_3\approx 0.02096,
\eea
The expectation value of the Ising order parameter is \cite{Fateev}
\be
\langle\sigma\rangle\approx 1.07496\ h^\frac{1}{15},
\ee
which enables us in principle to solve the self-consistency equation
\r{SCh}. The important point is that most of the spectral weight is
located in the coherent modes corresponding to the two lightest
particles. The ratio of weights between them is 
\be
\frac{(Z_1/m_1)}{(Z_2/m_2)}\approx 5.79427.
\ee
The region $m_0\approx 0$ limit was studied by form factor
perturbation theory in Ref.[\onlinecite{Mussardo}].
%%%%%%%%%%%%%%%%%%%%%%%%%%%%%%%%%%%%%%%%%%%%%%%%%%%%%%%%%%
\subsubsection{Qualitative Behaviour in the general case}
\begin{figure}[ht]
\noindent
\epsfxsize=0.45\textwidth
\epsfbox{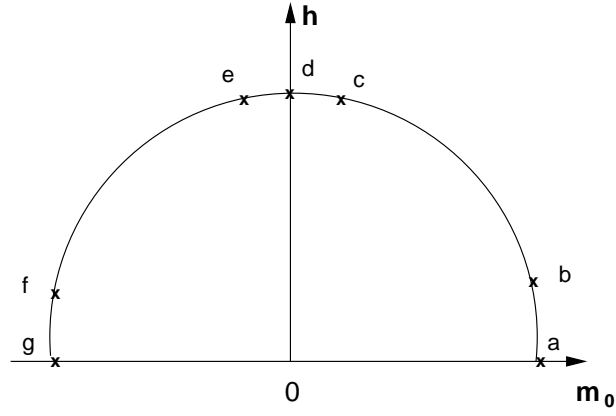}
\caption{Path in the $h-m_0$ plane of the transverse Ising model in a
magnetic field.
\label{fig:fig1}% 
} 
\end{figure}

%%%%%%%%%%%%%%%%%%%%%%%%%%%%%%%%%%%%%%%%%%%%%%%%%%%%%%%%%%
For general values of $\chi$ the qualitative behaviour of
$\chi^E_{\sigma\sigma}(\bar\omega,q)$ is known and may be
conveniently summarized \cite{McCoyWu1,Mussardo} by considering the
evolution of $\chi^E_{\sigma\sigma}$ with $\chi$ along a path in the
$m_0-h$ plane as shown in Fig. \ref{fig:fig1}. 

\begin{figure}[ht]
\noindent
\epsfxsize=0.45\textwidth
\epsfbox{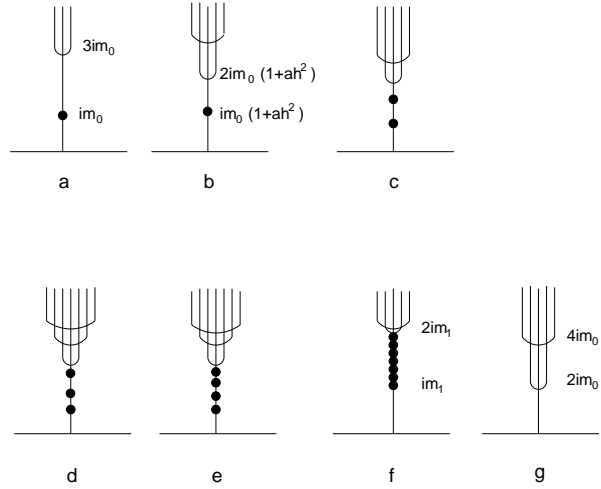}
\caption{Structure of poles and branch cuts of
$\chi^E_{\sigma\sigma}(\bar\omega,q)$ at the points (a)-(g) in the
$h-m_0$ plane indicated in Fig.\ref{fig:fig1} ($m_1=m_0[2+(\tilde{h}\lambda_1)^\frac{2}{3}]$).
\label{fig:fig2}% 
} 
\end{figure}

In Fig.\ref{fig:fig2} we show the analytic structure of
$\chi^E_{\sigma\sigma}$ as a function of
$s=\sqrt{\bar{\omega}^2+v^2q^2}$ for various locations along
the path set out in Fig.\ref{fig:fig1}. For example, point (a)
corresponds to the disordered phase of the off-critical Ising model,
where in Euclidean space there is a single-particle pole at $s=im_0$
and a 3-particle branch cut along the positive imaginary axis starting
at $s=3im_0$. Point (b) shows the small shift in the position of the
pole and the emergence of a 2-particle branchcut
\cite{McCoyWu1}. Points (c)-(e) describe the vicinity of the Ising
model in a magnetic field; there are several single-particle poles
below a 2-particle branchcut and the number of these poles increases
as we move along the path. Finally, points (f)-(g) describe the
breakup of the 2-particle branchcut mentioned above.

An important point is that for $m_0<0$, which corresponds to the
ordered phase of the Ising model for $h=0$, the general effect of the
magnetic field is to make the dynamical susceptibility look more
coherent in these sense that the low-energy regime is dominated by
single-particle poles. In particular, a weak interchain coupling in
the ordered phase close to $H_c$ leads to a disintegration of the
2-particle scattering continuum that dominates the dynamical structure
factor \r{DSFhigh} and the formation of a series of single-particle
poles. 

%%%%%%%%%%%%%%%%%%%%%%%%%%%%%%%%%%%%%%%%%%%%%%%
\subsection{Beyond Mean-Field: RPA}
%%%%%%%%%%%%%%%%%%%%%%%%%%%%%%%%%%%%%%%%%%%%%%%
It is straightforward to go beyond the mean-field approximation by
resumming all diagrams in the interchain coupling that do not involve
loops. This leads to the RPA expression for the dynamical
susceptibility \cite{RPA}
\bea
\chi^{xx}(\omega,q,{\bf k})&=&
\frac{\chi^{xx}(\omega,q)}
{1-2J_\perp({\bf k})\chi^{xx}(\omega,q)}\ ,
\eea
where $J_\perp({\bf k})$ is the Fourier transform of the interchain
coupling and 
\be
\chi^{xx}(\omega,q)=\frac{C^2}{a_0}\left[\frac{a_0}{v}\right]^\frac{1}{4}
\chi_{\sigma\sigma}(\omega,q).
\ee
In our notations 
\be
J_\perp({\bf    k})=J_\perp[\cos(k_ya_0)+\cos(k_za_0)]
\ee
for a simple cubic lattice. It was shown in Ref.[\onlinecite{CT}] that
in the vicinity of points (c)-(e) of Fig.\ref{fig:fig1} the RPA
leads only to slight changes of the mean-field results. More
precisely, single-particle excitations corresponding to poles at $s=m$
(with residue $Z/2m$) in the 1D susceptibility $\chi^{xx}(\omega,q)$
acquire a transverse dispersion $ZJ_\perp({\bf k})$ in the RPA.

%%%%%%%%%%%%%%%%%%%%%%%%%%%%%%%%%
\section{Summary and Discussion}
\label{summary}
%%%%%%%%%%%%%%%%%%%%%%%%%%%%%%%%%
We have studied the spectrum and dynamical spin correlations for
Haldane-gap systems in the presence of a magnetic field. We have
paid particular attention to the role played by the crystal field
anisotropies present in materials like NDMAP. We have concentrated on
the case where the magnetic field is applied along the same same direction
as the largest single-ion anisotropy $D$ (which we identify with the
z-direction in spin space). Generalizations of our results to other
cases is straightforward. Our main results are as follows:
\begin{itemize}
\item{} At low fields there are three coherent modes
$M_-$, $M_+$, $M_3$. Their respective gaps
$\Delta_-(H)<\Delta_+(H)<\Delta_3(H)$ are field-dependent. 
\item{} Above a critical field $H_{\rm d}$, $M_2$ develops a finite
lifetime {\sl via} the decay process $M_+\rightarrow M_-M_-M_-$. $M_-$
and $M_3$ retain infinite lifetimes.
\item{} At a critical field $H_c>H_{\rm d}$ the gap $\Delta_-(H)$
vanishes. In the vicinity of $H_c$ the low energy degrees of freedom
are described by an (off-critical) Ising model and the dynamical
structure factor is calculated by exact methods. For $H>H_c$ the
dynamical structure factor is dominated by an incoherent two-particle
scattering continuum above a finite-energy threshold. 
\item{} At fields sufficiently above $H_c$ the low-energy degrees of
freedom are described by a sine-Gordon model. $S^{\rm
yy}(\omega,\frac{\pi}{a_0}+q)$ is dominated by a coherent
single-particle bound state with a spectral gap below a two-particle
scattering continuum. The most pronounced feature in 
$S^{\rm xx}(\omega,\frac{\pi}{a_0}+q)$ is an incoherent
two-particle scattering continuum above a finite-energy threshold.
\item{} The effects of interchain coupling are most pronounced in the
vicinity of $H_c$. Taking it into account in a mean-field fashion
leads to a purely one-dimensional effective description at low
energies in terms of an {\sl Ising model in a magnetic field}. This
suggests that Haldane-gap materials with single-ion anisotropies in a
magnetic field may constitute a realization of this very interesting
theory. Within the mean-field description the main effect of
interchain coupling is to generate coherent single-particle modes from
the incoherent scattering continua. As a result the dynamical
structure factor will appear more ``coherent''.
\end{itemize}
Our findings shed some light on the question why recent inelastic
neutron scattering experiments on NDMAP \cite{az2,az3} have failed to
find any evidence of scattering continua in the high-field phase. At
large fields these are suppressed through bound-state formation,
whereas in the vicinity of the critical field $H_c$ the coupling
between chains effects similar shifts of spectral weight to
single-particle modes. It would be interesting to investigate some of
our predictions experimentally. In particular we hope that it may be
possible to
\begin{itemize}
\item[1.] address the issue of the finite lifetime of $M_2$ above
the critical field $H_{\rm d}$. 
\item[2.] disentangle the ${\rm xx}$  and ${\rm yy}$ components of the
structure factor at high fields. According to our predictions the
${\rm xx}$-component will remain incoherent up to fairly large fields
so that a scattering continuum may be observable.
\end{itemize}
%%%%%%%%%%%%%%%%%
\acknowledgments
%%%%%%%%%%%%%%%%%
We thank R.M. Konik, A.A. Nersesyan, S. Shapiro, A.M. Tsvelik,
I. Zaliznyak and A. Zheludev for important discussions. 
We acknowledge support by the U.S. Department of Energy under Contract
No DE-AC02-98 CH10886 (FE), the EPSRC under grant GR/R83712/01 (FE),
the Theory Institute for Strongly Correlated and Complex Systems at
BNL (IA), where much of this research was carried out, the Canadian
Institute for Advanced Research (IA) and NSERC of Canada (IA).

\appendix
%%%%%%%%%%%%%%%%%%%%%%%%%%%%%%%%%%%%%%%%%%%%%%%%%%%%%%%%%%%%%%%%%%%%%
\section{Low-Field Phase in the absence of crystal field anisotropy}
\label{app:U1}
%%%%%%%%%%%%%%%%%%%%%%%%%%%%%%%%%%%%%%%%%%%%%%%%%%%%%%%%%%%%%%%%%%%%%

In the absence of a crystal field anisotropy the Hamiltonian is
\be
{\cal H}(h)=J\sum_n {\bf S}_n\cdot {\bf S}_{n+1}-h{\bf S}^z.
\label{Hisoh}
\ee
The Heisenberg equations of motion read
\bea
\frac{d}{dt}S^\pm_n(t)&=&i[{\cal H}(0),S^\pm_n]\mp ih\ S^\pm_n\ ,\nn
\frac{d}{dt}S^z_n(t)&=&i[{\cal H}(0),S^z_n]\ .
\label{HeisenbergEOM}
\eea
Equations \r{HeisenbergEOM} permit us to express the dynamical
susceptibilities for $h\neq 0$ in terms of the ones in zero field
\bea
\chi^{+-}(\omega,q,h)&=&\chi^{+-}(\omega-h,q,0)\ ,\nn
\chi^{-+}(\omega,q,h)&=&\chi^{-+}(\omega+h,q,0)\ ,\nn
\chi^{zz}(\omega,q,h)&=&\chi^{zz}(\omega,q,0).
\eea

This implies is that the structure factors in a field are simply
the same as in zero field apart from constant shifts in energy. The
leading contributions to the dynamical susceptibilities in zero field
have been calculated in the framework of the O(3) nonlinear-sigma
model approximation to the isotropic spin-1 Heisenberg chain in Refs
[\onlinecite{ha,fab}]. In follows from these results that the
three-particle contributions are very small. We note 
that the threshold of the $M_+M_+M_-$ three-particle continuum 
(two $S^z=1$ magnons and one $S^z=-1$ magnon) is at $3\Delta-h$,
i.e. for any $H<H_c$ it still is very slightly higher in energy than
the highest energy $M_-$ magnon mode. Hence all three magnon 
modes remain ``sharp'' for all $H<H_c$.

We note that analogous considerations apply in the presence of a
single-ion anisotropy in z-direction only ($E=0$ in \r{heisenberg} and
a magnetic field along the z-direction). 
The dynamical susceptibilities for finite fields can then be expressed
in terms of the zero field susceptibilities through the equation of
motions for the spin operators. This is quite useful for the Majorana
fermion model, where the staggered components of the spin operators
are expressed in terms of Ising order and disorder operators. The
latter transform nontrivially under the Bogoliubov transformation used
to diagonalize the Hamiltonian for nonzero magnetic fields.

%%%%%%%%%%%%%%%%%%%%%%%%%%%%%%%%%%%%%%%%%%%%%%%%%%%%%%%%%%%%%%%%%%%%%
\section{Spectral representation of Correlation Functions}
\label{app:spectral}
%%%%%%%%%%%%%%%%%%%%%%%%%%%%%%%%%%%%%%%%%%%%%%%%%%%%%%%%%%%%%%%%%%%%%
In this appendix we collect useful formulas for spectral
representations of correlation functions in massive, integrable,
relativistic quantum field theories. 

We parametrize energy and momentum of single particle states
in terms of a rapidity variable $\theta$
\be
E_\eps(\theta)=\Delta_\eps\cosh\theta\ ,\quad
P_\eps(\theta)=\frac{\Delta_\eps}{v}\sinh\theta\ . 
\ee
Here the index $\eps$ labels the different types of particles and
$\Delta_\eps$ are the corresponding spectral gaps.
A scattering state of $N$ particles with rapidities $\{\theta_j\}$ and
indices $\{\eps_j\}$ is denoted by 
\be
|\theta_1,\theta_2,\ldots,\theta_N\rangle_{\eps_1,\eps_2,\ldots \eps_N}\
.
\label{basis}
\ee
Its energy and momentum are
\be
E(\{\theta_j\})=\sum_{j=1}^N\Delta_{\eps_j}\cosh\theta_j\ ,\
P(\{\theta_j\})=\sum_{j=1}^N\frac{\Delta_{\eps_j}}{v}\sinh\theta_j\ .
\ee
A basis of states is most easily constructed in terms of the generators
of the so-called Faddeev-Zamolodchikov algebra
\begin{eqnarray}
{Z}^{\epsilon_1}(\theta_1){Z}^{\epsilon_2}(\theta_2) &=&
{\bf S}^{\epsilon_1,\epsilon_2}_{\epsilon_1',\epsilon_2'}(\theta_1 -
\theta_2){Z}^{\epsilon_2'}(\theta_2){Z}^{\epsilon_1'}(\theta_1),\nn
{Z}_{\epsilon_1}^\dagger(\theta_1)Z_{\epsilon_2}^\dagger(\theta_2) &=&
Z_{\epsilon_2'}^\dagger(\theta_2){ Z}_{\epsilon_1'}^\dagger
(\theta_1){\bf S}_{\epsilon_1,\epsilon_2}^{\epsilon_1',\epsilon_2'}(\theta_1 -
\theta_2),\nn
Z^{\epsilon_1}(\theta_1)Z_{\epsilon_2}^\dagger(\theta_2) 
&=&Z_{\epsilon_2'}
^\dagger(\theta_2)
{\bf S}_{\epsilon_2,\epsilon_1'}^{\epsilon_2',\epsilon_1}
(\theta_2-\theta_1)Z^{\epsilon_1'}(\theta_1)\nn
&&+2 \pi\ \delta_{\epsilon_2}^{\epsilon_1} 
\delta (\theta_1-\theta_2).
\label{faza}
\end{eqnarray}
Here ${\bf  S}^{\epsilon_1,\epsilon_2}_{\epsilon_1',\epsilon_2'}(\theta)$ is 
the factorizable two-particle scattering matrix of the integrable
quantum field theory. Using the ZF operators a Fock space of states
can be constructed as follows. The vacuum is defined by
\begin{equation}
Z_{\varepsilon_i}(\theta) |0\rangle=0 \ .
\end{equation}
Multiparticle states are then obtained by acting with strings of
creation operators $Z_\epsilon^\dagger(\theta)$ on the vacuum
\begin{equation}
|\theta_n\ldots\theta_1\rangle_{\epsilon_n\ldots\epsilon_1} = 
Z^\dagger_{\epsilon_n}(\theta_n)\ldots
Z^\dagger_{\epsilon_1}(\theta_1)|0\rangle . 
\end{equation} 

The resolution of the identity in the normalization implied by
\r{faza} is given by
\begin{widetext}
\bea
1=\sum_{n=0}^\infty\frac{1}{n!}\
\sum_{\{\eps_j\}}
\int_{-\infty}^\infty \prod_{j=1}^n\frac{d\theta_j}{2\pi}
\ |\theta_n,\ldots,\theta_1\rangle_{\eps_n,\ldots \eps_1}\
^{\eps_1,\ldots \eps_n}\langle\theta_1,\ldots,\theta_n|\ .
\eea
The two point function of some operator ${\cal O}$ can now be
expressed in the spectral representation as
\bea
\langle{\cal O}^\dagger(t,x){\cal O}(0,0)\rangle=
\sum_{n=0}^\infty\frac{1}{n!}\sum_{\eps_j}
\int_{-\infty}^\infty\prod_{j=1}^n\frac{d\theta_j}{2\pi}
|{f}^{\cal O}_{\eps_1\ldots\eps_n}(\theta_1,\ldots,\theta_n)|^2
\exp\left(-itE(\{\theta_j\})+ixP(\{\theta_j\}\right)),
\label{2pt}
\eea
\end{widetext}
where the formfactors are given by
\be
\label{formf}
f_{\epsilon_1\ldots\epsilon_n}^{\cal O}(\theta_1\ldots\theta_n)
\equiv
\langle 0| {\cal
O}(0,0)|\theta_n\ldots\theta_1\rangle_{\epsilon_n\ldots\epsilon_1}.
\ee

%%%%%%%%%%%%%%%%%%%%%%%%%%%%%%%%%%%%%%%%%%%%%%%%%%%%%%%%%%%%%%%%%%%%%%%%%%%%
\section{Bound States in the Majorana Model}
\label{app:BoundStates}
%%%%%%%%%%%%%%%%%%%%%%%%%%%%%%%%%%%%%%%%%%%%%%%%%%%%%%%%%%%%%%%%%%%%%%%%%%%%

In this appendix we address the question of whether the
current-current interaction in the Majorana model leads to the
formation of bound states. For simplicity we consider only the SU(2)
symmetric case with Hamiltonian 
\bea
{\cal H}&=&\frac{i}{2}\int dx\sum_{a=1}^3 v[L_a\partial_xL_a- R_a\partial_xR_a]
-2mR_aL_a\nn
&&+g\int dx\sum_{a} J^aJ^a.
\label{Hsu2}
\eea
We aim to establish that bound states exist for any $g>0$, whereas
there are no bound states for $g<0$. We recall that the Majorana
fermion model arises from the spin-Heisenberg Hamiltonian with an
additional biquadratic term 
\bea
H_{\rm biquad}&=&J\sum_{n} {\bf S}_n\cdot {\bf S}_{n+1}- 
b\left({\bf S}_n\cdot{\bf S}_{n+1}\right)^2, 
\label{Hbiquad}
\eea
where $|b-1|\ll 1$. For $b>1$ the model is in a dimerized phase
whereas $b<1$ corresponds to a Haldane spin-liquid regime.
One may establish by using the expressions \r{decomposition} for
the spin operators that the case $b<1$ ($b>1$) corresponds to $g<0$
($g>0$). In order to determine whether the current-current interaction
leads to the formation of bound states, we first consider the limit of
a very anisotropic interaction
\bea
{\cal H}_{\rm ani}&=&\frac{i}{2}\int dx \sum_{a=1}^3 v[L_a\partial_xL_a-
R_a\partial_xR_a] -2mR_aL_a\nn
&&+g\int dx\ J^3J^3.
\eea
This case can be mapped onto a single massive Majorana fermion plus
the massive Thirring model by introducing complex Fermi fields by
\bea
R_1&=&\frac{\Psi_R+\Psi_R^\dagger}{\sqrt{2}}\ ,\quad
R_2=\frac{\Psi_R-\Psi_R^\dagger}{i\sqrt{2}}\ ,\nn
L_1&=&\frac{\Psi_L^\dagger-\Psi_L}{i\sqrt{2}}\ ,\quad
L_2=\frac{\Psi_L+\Psi_L^\dagger}{\sqrt{2}}\ .
\eea
The Hamiltonian density is rewritten as
${\cal H}_{\rm ani}={\cal H}_{\rm Maj}+{\cal H}_{\rm MTM}$, where
\bea
{\cal H}_{\rm Maj}&=&\frac{iv}{2}\int dx[L_3\partial_xL_3-
R_3\partial_xR_3 -\frac{2m}{v}R_3L_3],\nn
{\cal H}_{\rm
MTM}&=&-iv\int dx\left[\Psi^\dagger_R\partial_x\Psi_R-\Psi^\dagger_L 
\partial_x\Psi_L\right]\nn 
&+&\int dx\left[m[\Psi^\dagger_R\Psi_L+{\rm h.c.}]
+2g\Psi^\dagger_L\Psi_L\Psi^\dagger_R\Psi_R\right].\nn
\label{Hmtm}
\eea
In Eqn \r{Hmtm} we have dropped a term proportional to
$ \int dx [ \Psi^\dagger_R \Psi_R + \Psi^\dagger_L\Psi_L]$ as it commutes
with the Hamiltonian. It is well known that in the massive Thirring
model there are breather bound states for $g>0$, but no bound states
exist for $g<0$, see e.g. Refs[\onlinecite{MTM}]. 

A different approach is to use large-N methods. If we consider
$N$ species of Majorana fermions rather than three, we may decouple
the interaction through a bosonic Hubbard-Stratonovich field
$\sigma$. For even $N$ and $g>0$ the problems maps onto the O(N/2) massive
Gross-Neveu model, which is known to have bosonic bound states in the
large-N limit \cite{GN}.

%%%%%%%%%%%%%%%%%%%%%%%%%%%%%%%%%%%%%%%%%%%%%%%%%%%%%%%%%%%%%%%%%%
\section{Low-energy projections of the staggered magnetizations}
\label{operatorcontent}
%%%%%%%%%%%%%%%%%%%%%%%%%%%%%%%%%%%%%%%%%%%%%%%%%%%%%%%%%%%%%%%%%%
In this appendix we give arguments in favour of the identification
\r{ops} at low energies and in the vicinity of the Ising critical
point at $H=H_c$. We first consider the LG theory and then the
Majorana fermion model.

%%%%%%%%%%%%%%%%%%%%%%%%%%%%%%%%%%%
\subsection{Landau-Ginzburg Model}
%%%%%%%%%%%%%%%%%%%%%%%%%%%%%%%%%%%
It is instructive to examine the evolution of the amplitudes
$A_{a\alpha}$ entering the mode expansions \r{LGmodes} of the scalar
fields $\varphi_{a}$ as the magnetic field is increased. 
We recall that the critical field is $H_c=\Delta_1$. In the vicinity
of $H_c$ we parametrize
\be
H=\Delta_1-\delta\ ,\quad \delta >0.
\ee
As we are interested only in low energies we may restrict our
attention to the ``$-$'' modes. From \r{apm} we obtain the following
expansions in $\delta$ 
\bea
\left(\omega_-(0)\right)^2&\approx&
\frac{2\Delta_1(\Delta_2^2-\Delta_1^2)}{3\Delta_1^2+\Delta_2^2}\delta
\ ,\nn
\left[\frac{A_{2-}(0)}{A_{1-}(0)}\right]^2&=&
\frac{(7\Delta_1^2+\Delta_2^2)\delta}{2\Delta_1(3\Delta_1^2+\Delta_2^2)}\ .
\label{apm3}
\eea
Eqns \r{apm3} imply that close to $H_c$ we have
\be
A_{2-}(0)\propto \omega_-(0)A_{1-}(0)\ .
\label{apm4}
\ee
As we have seen before, close to $H_c$ the x-component of
the staggered magnetization $\varphi_1$ couples to $M_-$ with a finite
amplitude $A_{1-}(0)$ given by \r{apm2}. Furthermore we have the
identification \r{ops}
\bea
\varphi_1\propto\sigma\ ,
\label{ops3}
\eea
where $\sigma$ is the Ising order parameter field. Eqns \r{apm4} and
\r{ops3} together suggest that
\be
\varphi_2\propto\partial_t\sigma\ .
\ee
This claim may be substantiated further in the limit where one of the
zero field gaps is much smaller than the other,
i.e. $\Delta_1\ll\Delta_2$. As we are interested in energies that are
small compared to $\Delta_2$, we may ``integrate out'' the high-energy
degrees of freedom corresponding to $\varphi_2$ in the path integral
expression for the staggered magnetization $n^y$. Because
$H_c=\Delta_1$ is small, we furthermore may take the magnetic field
into account perturbatively. The staggered magnetization in $y$
direction is   
\be
n^y(t,x)=\varphi_2(t,x)\ .
\ee
Averaging $n^y(t,x)$ over $\varphi_2$, we obtain 
\bea
\langle n^y(t,x)\rangle_2&=&\frac{1}{Z}\int{\cal D}\varphi_2\
\varphi_2(t,x)\nn
&&\times e^{iS_2-2i(H/v)\int dt_1 dx_1[\varphi_2\partial_{t_1}\varphi_1]},
\eea
where
\bea
S_2=\int dt dx\
\left[\frac{1}{2v}\left(\partial_t\varphi_2\right)^2
-\frac{v}{2}\left(\partial_x\varphi_2\right)^2
-\frac{\Delta_2^2}{2v}\varphi^2_2\right].
\eea
The leading contribution occurs in first order in the magnetic field
\bea
&&\langle n^y(t,x)\rangle_2\nn
&\approx&-\frac{2iH}{v}\int dx_1dt_1\ \langle T \varphi_2(t,x)\varphi_2(t_1,x_1)\rangle_2\
\partial_t\varphi_1(t_1,x_1)\nn
&=&\frac{2H}{v}\int dx_1dt_1\ G_2(t-t_1,x-x_1)\ \partial_t\varphi_1(t_1,x_1)\nn
&\approx&\frac{2H}{v}\left[\int dx'_1dt'_1\ G_2(t'_1,x'_1)\right]\ 
 \partial_t\varphi_1(t,x)\ .
\eea
In the last line we have used that the leading contribution to the
integral comes from the region $t_1\approx t$, $x_1\approx x$.
This shows that the mixing induced by the magnetic field generates a
contribution to $n^y(t,x)$ proportional to $\partial_t\varphi_1(t,x)$
at low energies.

%%%%%%%%%%%%%%%%%%%%%%%%%%%%%%%%%%%
\subsection{Majorana Fermion Model}
%%%%%%%%%%%%%%%%%%%%%%%%%%%%%%%%%%%
Analogous calculations can be performed in the framework of the
Majorana fermion model in the case $g_a=0$, i.e. in the absence of the
current-current interactions. In particular, let us consider the case
where one of the zero field gaps is much smaller than the other
\be
\Delta_1\ll\Delta_2\ .
\ee
As the critical field $H_c=\sqrt{\Delta_1\Delta_2}$ is much
smaller than $\Delta_2$ we may treat the magnetic field term
perturbatively. We may derive an effective action for $R_1$,
$L_1$ only by integrating out $R_2$ and $L_2$ (we recall that the
third Majorana decouples in the absence of interactions)
\bea
S_{\rm eff}&\approx&S_1-\frac{1}{2}\langle S_H^2\rangle_2\ ,\nn
S_H&=&iH\int dx\ d\tau\left[L_1L_2+R_1R_2\right]\ ,
\eea
where $\langle\rangle_2$ denotes the expectation value with respect to
the second Majorana fermion and
\bea
S_1&=&\int d\tau\ dx\left[R_1\partial_-R_1+ L_1\partial_+L_1
  -i\Delta_1R_1L_1\right],\nn
\partial_\pm&=&\frac{\partial_\tau\pm iv\partial_x}{2}\ .
\eea
The Matsubara Green's functions are defined as e.g.
\be
G_{RR}(\tau,x)=-\langle T_\tau\ R(\tau,x)\ R(0)\rangle\ .
\ee
Their Fourier transforms are
\bea
G_{R_2R_2}(\omega,q)&=&-\frac{i\omega+vq}{\omega^2+v^2q^2+\Delta_2^2}\ ,\nn
G_{L_2L_2}(\omega,q)&=&-\frac{i\omega-vq}{\omega^2+v^2q^2+\Delta_2^2}\
,\nn
G_{R_2L_2}(\omega,q)&=&+\frac{i\Delta_2}{\omega^2+v^2q^2+\Delta_2^2}\ .
\eea
A straightforward calculation then gives
\bea
{\cal L}_{\rm eff}&=R_1\partial_-R_1+ L_1\partial_+L_1
  -i[\Delta_1-\frac{H^2}{\Delta_2}]R_1L_1\ .
\label{Leff}
\eea
In other words, integrating out the second Majorana leads to a
renormalization of the mass of the first Majorana. We note that the
dispersion relation that follows from \r{Leff} agrees with the
expansion of $\omega_-(q)$ \r{opm} in the case $H\ll\Delta_2$ as it
must. 
What we have shown is that in the case $\Delta_1\ll\Delta_2$ it is
simply the first Majorana that becomes critical at $H_c$. The
staggered magnetization in $x$-direction is expressed at low energies
by averaging \r{staggered} with respect to the second and third
Majoranas, which gives
\bea
n^x(x)\propto\sigma^1(x)\langle\mu^2(x)\rangle\langle\mu^3(x)\rangle. 
\eea
The determination of the operator content of $n^y(x)$ at low energies
is significantly more involved. In order to obtain the low-energy
projection,  we need to average with respect to the second and third
Majoranas
\bea
&& n^y(0,0)\approx -iH
\langle\mu_3\rangle\ \nn
&&\times \int d\tau\ dx
\Bigl\lbrace\langle L_2(\tau,x)\ \sigma_2(0)\rangle_2\ L_1(\tau,x)\ \mu_1(0)\nn
&&\qquad\quad
+\langle R_2(\tau,x)\ \sigma_2(0)\rangle_2\ R_1(\tau,x)\ \mu_1(0)\Bigr\rbrace\ .
\label{ny}
\eea
The expectation values $\langle\rangle_2$ with respect to the second
Majorana can be evaluated by the form factor bootstrap approach by
utilizing the results of Refs[\onlinecite{barouch,yurov,cardy}]. We obtain
\bea
\langle L_2(\tau,x)\ \sigma_2(0)\rangle_2&\approx&D\ e^{i\pi/4}
\frac{1}{\sqrt{v\tau+ix}}e^{-mr}\ ,\nn
\langle R_2(\tau,x)\ \sigma_2(0)\rangle_2&\approx&D\ e^{-i\pi/4}
\frac{1}{\sqrt{v\tau-ix}}e^{-mr}\ ,
\label{LRsigma}
\eea
where $D=m^{1/8}(4\pi)^{-1/2}2^{1/12}e^{-1/8}A^{3/2}$ and 
$r^2=\tau^2+x^2/v^2$. Using these results in \r{ny} we see that
the integral is dominated with exponential accuracy by the region
$\tau\approx 0$, $x\approx 0$. Hence the operator content of $n^y$ is
determined by the fusion of the disorder operator $\mu_1$ with the
left and right moving fermions $L_1,R_1$. The relevant operator
product expansions can be worked out following
Ref.[\onlinecite{Fonseca}] 
\bea
L(\tau,x) \mu(0)\approx \frac{\gamma}{\sqrt{z}}\sigma(0)
-\frac{m\gamma}{v}\sqrt{{\bar z}}\sigma(0)
+\frac{4\gamma}{v}\sqrt{z}\partial_-\sigma(0),\nn
R(\tau,x) \mu(0)\approx \frac{\bar\gamma}{\sqrt{\bar z}}\sigma(0)
-\frac{m\bar\gamma}{v}\sqrt{z}\sigma(0)
+\frac{4\bar\gamma}{v}\sqrt{\bar z}\partial_+\sigma(0),\nn
\label{OPE}
\eea
where $z=v\tau+ix$, $\gamma=\exp(-i\pi/4)/\sqrt{4\pi}$ and
$\partial_\mp=\frac{1}{2}(\partial_\tau\mp iv\partial_x)$.
Combining \r{LRsigma} with \r{OPE} we obtain the desired result
\be
n^y\propto \partial_\tau\sigma\ .
\ee

%%%%%%%%%%%%%%%%%%%%%%%%%%%%%%%%%%%%%%%%%%%%%%%%%%%%%%%%%%%%%%%%%%%%%%
\section{Derivation of the sine-Gordon model in the high-field phase
for weak anisotropy}
\label{sineG}
%%%%%%%%%%%%%%%%%%%%%%%%%%%%%%%%%%%%%%%%%%%%%%%%%%%%%%%%%%%%%%%%%%%%%%

In this appendix we show how the Sine-Gordon Hamiltonian emerges as
the low-energy effective theory at $H>H_c$ in the small anisotropy
limit $\Delta_2-\Delta_1\ll H-H_c$. We first present a derivation in
the framework of the nonlinear sigma model and then within the
Majorana fermion model.  
%%%%%%%%%%%%%%%%%%%%%%%%%%%%%%%%%%%%%
\subsection{Nonlinear sigma model}
%%%%%%%%%%%%%%%%%%%%%%%%%%%%%%%%%%%%%
The isotropic spin-S Heisenberg chain in a magnetic field can be
mapped onto the O(3) nonlinear sigma model in the continuum limit.
Exploiting the integrability of the nonlinear sigma model
it was shown in Ref. [\onlinecite{KF}] that for $H>\Delta$ the
low-energy regime is described in terms of a free boson
\bea
{\cal H}=\frac{\tilde{v}}{16\pi}\int dx\ \left[(\partial_x\Phi)^2
+(\partial_x\Theta)^2\right].
\eea
Here $\Theta$ is the field dual to $\Phi$ and fulfils
\bea
\tilde{v}\partial_x\Theta=-i\partial_\tau\Phi\ ,\quad
\partial_\tau\Theta=i\tilde{v}\partial_x\Phi\ .
\eea
The low-energy behaviour of spin correlations follows from the
correspondence 
\bea
S_n^\pm&\simeq& (-1)^n\ A\exp(\pm i\frac{\beta}{2}\Theta)\ .
%,\nn
%S_n^z&\simeq&
%\frac{a_0}{4\pi\beta}\partial_x\Phi+B\cos\left[\frac{1}{2\beta}\Phi-2\pi
%Mx\right]. 
\eea
The parameters $\tilde{v}$ and $\beta$ were calculated 
in Ref. [\onlinecite{KF}]. Adding a very small crystal field
anisotropy to the Hamiltonian
\be
E\sum_j[(S^x_j)^2-(S^y_j)^2],
\ee
generates a term proportional to
\be
\int dx \cos(\beta\Theta)\ .
\ee
The resulting theory is the sine-Gordon model \r{SGM}.
The term $D\sum_j(S^z_j)^2$ merely leads to a small change in $\beta$
which we ignore here.
%%%%%%%%%%%%%%%%%%%%%%%%%%%%%%%%%%%%%
\subsection{Majorana fermion model}
%%%%%%%%%%%%%%%%%%%%%%%%%%%%%%%%%%%%%
Our starting point is the
Hamiltonian \r{Heff} describing the two Majorana fermions that couple to
the magnetic field in the limit $\Delta_2=\Delta_1=m$, i.e. vanishing
gap anisotropy $\Delta=0$. Using
\be
\Psi_{R,L}=\int_{-\infty}^\infty\frac{dk}{2\pi} \ e^{ikx} c_{R,L}(k)\ ,
\ee
we may express the Hamiltonian as
\bea
{\cal H}_{\rm 12}\biggl|_{\Delta=0}&=&
\int_{-\infty}^\infty\frac{dk}{2\pi}
(c^\dagger_R,c^\dagger_L)\ M\
\pmatrix{c_R \cr c_L}
\eea
where
\be
M=\left(
\begin{array}{cc}
vk+H & -im\\
im & -vk+H\\
\end{array}\right).
\ee
Now we may carry out a Bogoliubov transformation
\be
\pmatrix{a_k\cr b_k\cr}=
\left(\begin{array}{cc}
\cos(\varphi_k) & -i\sin(\varphi_k) \\
-i\sin(\varphi_k) & \cos(\varphi_k) \\
\end{array}\right)
\pmatrix{c_R(k)\cr c_L(k)\cr}
\ee
with
\be
\tan(2\varphi_k)=\frac{m}{vk}
\ee
to diagonalize the Hamiltonian. We find
\bea
{\cal H}_{\rm 12}\biggl|_{\Delta=0}&=&
\int\frac{dk}{2\pi}\Bigl[(H+{\rm sgn}(k)\sqrt{m^2+v^2k^2})\ a^\dagger_ka_k\nn
&&+(H-{\rm sgn}(k)\sqrt{m^2+v^2k^2})\ b^\dagger_kb_k\Bigr].
\label{modes2}
\eea
Introducing fermions $c$ and $d$ by
\bea
c(k)&=&a_k\theta(k)+b_k\theta(-k)\ ,\nn
d(k)&=&b_k\theta(k)+a_k\theta(-k)\ ,
\eea
we may express the Hamiltonian \r{modes2} as
\bea
{\cal H}_{\rm 12}\biggl|_{\Delta=0}&=&
\int\frac{dk}{2\pi}\Bigl[(H+\sqrt{m^2+v^2k^2})\ c^\dagger(k)c(k)\nn
&&+(H-\sqrt{m^2+v^2k^2})\ d^\dagger(k)d(k)\Bigr].
\eea
The low-energy modes occur in the lower band in the vicinity of $\pm
k_F=\pm\sqrt{(H^2-m^2)/v^2}$. They can be combined into left and
right moving Fermi fields by
\be
d(x)=\exp(-ik_Fx)R(x)+\exp(ik_Fx)L(x)\ .
\label{dRL}
\ee
The low-energy effective Hamiltonian is then
\bea
{\cal H}'=i\tilde{v}\int dx\ \left[L^\dagger\partial_xL
-R^\dagger\partial_xR\right],
\eea
where $\tilde{v}=v^2k_F/H$. We now bosonize the low-energy Hamiltonian
using
\bea
R^\dagger(x)&\sim&
\frac{1}{\sqrt{2\pi}}\exp\left(i\frac{\Phi(x)+\Theta(x)}{2\sqrt{2}}
\right),\nn
L^\dagger(x)&\sim&
\frac{1}{\sqrt{2\pi}}\exp\left(-i\frac{\Phi(x)-\Theta(x)}{2\sqrt{2}}\right).
\eea
Here $\varphi$ and $\bar{\varphi}$ are chiral Bose fields fulfilling
the commutation relations $[\varphi(x),\bar{\varphi}(y)]=2\pi i$. In
terms of the canonical Bose field $\Phi=\varphi+\bar{\varphi}$ and the
dual field $\Theta=\varphi-\bar{\varphi}$ we find
\bea
{\cal H}'=\frac{\tilde{v}}{16\pi}\int dx\ \left[(\partial_x\Phi)^2
+(\partial_x\Theta)^2\right].
\label{hprime}
\eea
The high-energy cutoff in this construction is given by the depth of
the Fermi sea in the lower band of \r{modes2}, which is $H-m=H-H_c$. 
So far we have neglected the gap anisotropy, i.e. the term
\be
{\cal H}_{\rm pair}
=i\Delta\left[\Psi^\dagger_R\Psi^\dagger_L-h.c.\right]
\ee
in the Hamiltonian \r{Heff}. In the next step we take
it into account under the assumption that it $\Delta$ is small
compared to the cutoff $H-H_c$. In this limit ${\cal H}_{\rm pair}$
is expressed in terms of the modes as 
\bea
{\cal H}_{\rm pair}=
i\Delta\int_{-\infty}^\infty\frac{dk}{2\pi}
\left[c^\dagger_R(k)c^\dagger_L(-k)-c_L(-k)c_R(k)\right].\nn
\eea
After the Bogoliubov transformation this becomes
\bea
&&-i\Delta\int_0^\infty\frac{dk}{2\pi}
\cos(2\varphi_k)\left[d(k)d(-k)-c(k)c(-k)-{\rm h.c.}\right]\nn
&&+{\rm mixed\  terms}.
\eea
Dropping the ``high-energy'' filled band as well as the mixed terms (they contribute in
higher orders of $\Delta/(H-H_c)$) we have
\bea
{\cal H}'_{\rm pair}\simeq
-i\Delta\int_0^\infty\frac{dk}{2\pi}
\frac{vk}{\sqrt{m^2+v^2k^2}}\left[d(k)d(-k)-{\rm h.c.}\right].\nn
\eea
Expanding around $\pm k_F$ this can be rewritten in terms of the left
and right moving fermions as
\bea
{\cal H}'_{\rm pair}\simeq
i\Delta\sqrt{1-\frac{m^2}{H^2}}\int dx\left[RL-L^\dagger R^\dagger\right].
\eea
Finally, bosonization gives
\bea
{\cal H}'_{\rm pair}\simeq
-\frac{\Delta}{\pi}\sqrt{1-\frac{m^2}{H^2}}
\int dx\ \cos\left(\frac{\Theta}{\sqrt{2}}\right).
\label{hprimepair}
\eea
By combining Eqns \r{hprime} and \r{hprimepair} we see that in the
absence of interactions the Majorana fermion model gives rise to a
sine-Gordon effective theory at low energies in the parameter regime
we have been discussing. The parameter $\beta$ in \r{SGM} takes the
special free-fermionic value $\beta=1/\sqrt{2}$. 

Interactions can be treated in a way analogous to the pairing term.
If we drop the interaction terms involving the third Majorana (which
we assume to have the largest gap) the interaction Hamiltonian reads
\bea
{\cal H}_{\rm int}=2g_3\int dx\ L_1L_2R_1R_2\ ,
\eea
where $g_3<0$.
Expressing this in terms of the complex fields $\Psi_{R,L}$, carrying
out a mode expansion and subsequent Bogoliubov transformation and finally
projecting to the low-energy band we obtain
\be
{\cal H}'_{\rm int}\simeq
2g_3\frac{v^2k_F^2}{m^2+v^2k_F^2}\int dx\ R^\dagger R L^\dagger L,
\ee
where $R$ and $L$ have been introduced in \r{dRL}. Bosonization then
gives 
\be
{\cal H}'_{\rm int}\simeq
\frac{g_3'\tilde{v}}{16\pi}\int dx\
\left[(\partial_x\Phi)^2-(\partial_x\Theta)^2\right], 
\ee
where $g_3'=\frac{g_3\tilde{v}}{\pi v^2}$. This term
may be combined with \r{hprime} by rescaling the scalar fields in the
standard way
\be
\Phi\longrightarrow \left[\frac{1-g_3'}{1+g_3'}\right]^\frac{1}{4}\Phi\ ,\quad
\Theta\longrightarrow \left[\frac{1+g_3'}{1-g_3'}\right]^\frac{1}{4}\Theta\ .
\ee
In terms of the rescaled fields the total Hamiltonian takes the form
of a sine-Gordon model \r{SGM} (with a slightly changed velocity)
%\bea
%&&{\cal H}'+{\cal H}'_{\rm int}+{\cal H}'_{\rm pair}\simeq
%\frac{\tilde{v}}{16\pi}\int dx\
%\left[(\partial_x\Phi)^2+(\partial_x\Theta)^2\right]\nn
%&&\qquad\qquad\qquad-\frac{\Delta}{\pi}\sqrt{1-\frac{m^2}{H^2}}
%\int dx\ \cos\left(\beta\Theta\right),
%\label{sgmapp}
%\eea
where 
\be
\beta=\left[\frac{1+g_3'}{1-g_3'}\right]^\frac{1}{4}\frac{1}{\sqrt{2}}<\frac{1}{\sqrt{2}}\ .
\ee
Hence the sine-Gordon model is in the attractive regime.

%%%%%%%%%%%%%%%%%%%%%%%%%%%%%%%%%%%%%%%%%%%%%%%%%%%%%%%%%%%%%%%%%%
\section{Correlation Amplitude in the Commensurate-Incommensurate 
Transition}
\label{app:corramp}
%%%%%%%%%%%%%%%%%%%%%%%%%%%%%%%%%%%%%%%%%%%%%%%%%%%%%%%%%%%%%%%%%%
Let us consider the LG model \r{LG} in the U(1) symmetric case
$\Delta_1=\Delta_2=\Delta_3=m$. For $H>H_c=m$ the low-energy degrees of
freedom are described by a Luttinger liquid, which can be derived by
means of Haldane's harmonic fluid approach \cite{haldane} as follows.
Forming a complex Bose field out of the two components of the LG field
that couple to the magnetic field
\be
\Psi_B=\frac{\varphi_1+i\varphi_2}{\sqrt{2}}\ ,
\ee
and then bosonizing using \cite{haldane}
\be
\Psi_B^\dagger\simeq \sqrt{\rho_0+a_0\Pi}
\left[\sum_{m\ {\rm even}}e^{im\Theta}\right] e^{i\Phi}\ ,
\label{bosonizationH}
\ee
one obtains, after rescaling the fields $\Phi$ and $\Theta$, a
Lagrangian density of the form \cite{ian2} 
\be
{\cal L}=\frac{1}{16\pi}\left[\frac{1}{\tilde{v}}
\left(\frac{\partial\Theta}{\partial t}\right)^2-
\tilde{v}\left(\frac{\partial\Theta}{\partial x}\right)^2\right].
\ee
Here $\tilde{v}=2v\sqrt{2\frac{H-m}{m}}$, 
$\rho_0$ is the (dimensionless) boson density, which
corresponds to the magnetization per site and $\Pi$ is the
momentum conjugate to $\Phi$. As the LG fields $\varphi_a$ are the
staggered components of the spin operators we conclude that
\bea
S_j^\pm\propto (-1)^j A \exp\left(\pm
\frac{i\beta\Theta}{2}\right)\ ,
\label{sjpm}
\eea
where $\beta$ depends on the magnetization and is related to the
parameter $\eta$ of Ref. [\onlinecite{ian2}] by $\beta^2=\eta$. 
By virtue of \r{bosonizationH} the amplitude $A$ is proportional to  
\be
A\propto 
\sqrt{\rho_0}a^{\frac{\beta^2}{2}}\ , 
\label{Ampl}
\ee
where $a$ is a short-distance cutoff and vertex operators are
normalized according to \r{VON}. The short-distance cutoff is
\be
a=\frac{a_0}{\rho_0}\ .
\label{SDC}
\ee
We note that the short-distance cutoff \r{SDC}  diverges as $H$
approaches $m$ as $\rho_0\to 0$ and \r{sjpm} describes the asymptotic
behaviour of spin correlation functions at distances much larger than
$a$. Combining \r{SDC} and \r{Ampl} we find that in general we have
\cite{haldane} 
\be
A\propto \rho_0^{\frac{1-\beta^2}{2}}
\ee
Let us now specialize to magnetic fields very close to the critical
field $H_c=m$ 
\be
H-m\ll m.
\ee
As shown in Ref. [\onlinecite{haldane}], the parameter $\beta$ tends to
$\frac{1}{\sqrt{2}}$, so that
\bea
S_j^\pm\propto (-1)^j A \exp\left(\pm
\frac{i\Theta}{2\sqrt{2}}\right)\ ,
\eea
where the density is given by \cite{ian2}
\be
\rho_0=\frac{a_0}{\pi v}\sqrt{2m(H-m)}.
\label{rho0}
\ee
Combining \r{rho0}, \r{SDC} and \r{Ampl} we obtain
\bea
A=A'a_0^\frac{1}{4}\
\left[\frac{H-H_c}{J}\right]^\frac{1}{8},
\label{aprime}
\eea
where $A'$ is a numerical, field independent constant and where we
have used that $\frac{v^2}{ma_0^2}\propto J$. The field dependence of
the correlation amplitude \r{aprime} is a universal feature of the
C-IC transition.

Let us apply these ideas to another example of the C-IC transition:
the spin-1/2 Heisenberg XXZ chain in a longitudinal magnetic field
\bea
{\cal H}_{\rm XXZ}&=&J\sum_n S^x_nS^x_{n+1}+S^y_nS^y_{n+1}+\delta
S^z_nS^z_{n+1}\nn
&& + H\sum_n S^z_n\ ,
\label{XXZ}
\eea
where $-1<\delta\leq 1$. The model \r{XXZ} has a phase transition from
a gapless, incommensurate Luttinger liquid phase to a gapped,
commensurate, spin-polarized phase at a critical value
\be
H_c=J(1+\delta).
\ee
Slightly below this transition, i.e.
\be
0<1-\frac{H}{H_c}\ll 1\ ,
\ee
the transverse correlation functions exhibit the following
large-distance asymptotics
\bea
\langle S^x_1(0)\ S^x_{R+1}(0)\sim (-1)^R
\frac{e^\frac{1}{2}2^\frac{11}{12}}{4A^6}
\frac{([H_c-H]/J)^\frac{1}{4}}{R^\frac{1}{2}},
\label{AXXZ}
\eea
where $A$ is Glaisher's constant \r{glaisher}. This result is obtained
as follows: the dependence on the magnetic field is universal and
given by \r{aprime}. The numerical coefficient is fixed by noting that
the numerical results of Ref.[\onlinecite{HiFu}] show that $A'$ is
independent of the value of the anisotropy $\delta$. Finally, we use
that the correlation amplitude has been calculated for the free
fermion case in Ref.[\onlinecite{Tracy}]. The result \r{AXXZ} is in
good agreement with the numerical results of
Ref.[\onlinecite{HiFu}] in the close proximity of the transition, as
was already noted in \cite{HiFu}.

%%%%%%%%%%%%%%%%%%%%%%%%%%%%%%%%%%%%%%%%%%%%%%%%%%%%%%%%%%%%%%%%%%5

\end{document}